# Deep Learning of Dynamic Subsurface Flow via Theory-guided Generative Adversarial Network

Tianhao He[a] and Dongxiao Zhang[b,c*]

[a]College of Engineering, Peking University, Beijing 100871, P. R. China
[b]School of Environmental Science and Engineering, Southern University of Science and Technology, Shenzhen 518055, P. R. China
[c]Intelligent Energy Laboratory, Peng Cheng Laboratory, Shenzhen 518000, P. R. China
[*]Corresponding author: E-mail address: zhangdx@sustech.edu.cn (Dongxiao Zhang)

**Abstract**

Generative adversarial network (GAN) has been shown to be useful in various applications, such as image recognition, text processing and scientific computing, due its strong ability to learn complex data distributions. In this study, a theory-guided generative adversarial network (TgGAN) is proposed to solve dynamic partial differential equations (PDEs). Different from standard GANs, the training term is no longer the true data and the generated data, but rather their residuals. In addition, such theories as governing equations, other physical constraints and engineering controls, are encoded into the loss function of the generator to ensure that the prediction does not only honor the training data, but also obey these theories. TgGAN is proposed for dynamic subsurface flow with heterogeneous model parameters, and the data at each time step are treated as a two-dimensional image. In this study, several numerical cases are introduced to test the performance of the TgGAN. Predicting the future response, label-free learning and learning from noisy data can be realized easily by the TgGAN model. The effects of the number of training data and the collocation points are also discussed. In order to improve the efficiency of TgGAN, the transfer learning algorithm is also employed. Numerical results demonstrate that the TgGAN model is robust and reliable for deep learning of dynamic PDEs.

## 1. Introduction

In recent years, various deep learning, algorithms such as artificial neural network (ANN), recurrent neural network (RNN) and convolutional neural network (CNN), have been rapidly developed and widely utilized in various disciplines. As a new type of neural network, proposed by Goodfellow et al. (2014), the generative adversarial network (GAN), can be used in modeling complex data distributions. GAN has achieved remarkable success in artificial intelligence (AI), including image recognition (Chen et al., 2016; Dupont et al., 2018; Pan et al., 2018), text processing (Liang et al., 2017; Yu et al., 2017), and scientific computing (Yang et al., 2020).

    Standard GANs can learn the distribution of target data only in data-driven form, which lack robustness when the database is not sufficiently large, or when the data of

the entire domain cannot be fully obtained. In the case of time-varying distributions, it is challenging for GANs to predict future responses. Furthermore, it is difficult for the generated results of standard GANs to resist the influence of noise, as it corrupts the true distribution.

In order to improve the performance of GANs, to make the generated data more consistent with physical laws, some researchers recently proposed to encode physical constraints into GANs. Physics-informed GANs (PI-GANs) can solve stochastic differential equations (SDEs) (Yang et al., 2020), after incorporating the governing physical laws into the architecture of GANs in the form of SDEs using automatic differentiation. Stinis et al. (2019) enforced constraints for interpolation and extrapolation to augment the efficiency of GANs. Yang and Perdikaris (2018) integrated the original GAN loss function with the physical constraint to cause the prediction to satisfy the governing laws. Yang et al. (2019) proposed conditional GANs (cGANs) by extending the loss function of generators. Wu et al. (2020) utilized GANs with statistical constraints to model chaotic dynamical systems. Zheng et al. (2019) used GANs to infer different heterogeneous fields simultaneously by incorporating the physical connections between them. Almost all of these works, however, were carried out around steady state data, whose distribution does not change with time.

In this work, we propose a theory-guided generative adversarial network (TgGAN) framework for dynamic problems, which enforces the theories into the loss function of generators in the form of soft constraints. Through switching the training target to the residuals between true data and generated data, the TgGAN can learn the mapping relation between the input and output, so as to solve the problem of data distributions changing with time. In the TgGAN, the generator is trained not only with available data at limited time steps, but also by honoring the physical principles and other theories at all time steps. As a consequence, the capabilities of interpolation, extrapolation, and generalization are greatly expanded. In this work, the proposed TgGAN framework is demonstrated with dynamic subsurface flow problems with heterogeneous parameters. Several two-dimensional subsurface flow problems with different scenarios, including training from noisy data and changing boundary conditions, are designed to test the performance of the TgGAN against the standard GAN. The effects of the number of training data and collocation points are also discussed, respectively. This work is inspired by and built upon the recent work of the theory-guided neural network (TgNN) (Wang et al., 2020). However, different from a single network in TgNN, which updates the network parameters by minimizing mean square error (MSE), the generator and the discriminator are trained in an adversarial manner in TgGAN. Moreover, instead of a fully connected neural network, the TgGAN makes use of CNNs.

The remainder of this paper proceeds as follows. In section 2, we introduce the governing equation, GAN, Wasserstein GAN with gradient penalty (WGAN-GP) and CNN, and present the architecture of TgGAN. In section 3, several two-dimensional cases are designed to test the performance of the TgGAN. Finally, the discussion and conclusions are provided in section 4.

## 2. Methodology

In this section, we first briefly introduce the governing equation with heterogeneous model parameters. Then, we elaborate on the standard GAN, WGAN-GP, and the basic framework of TgGAN. CNN is also briefly introduced. Finally, we outline the architecture of TgGAN and how to incorporate the governing equation into the TgGAN.

### 2.1 Governing equation with heterogeneous parameter

The parameter fields of subsurface flow problems are usually heterogeneous. The governing equation of flow in a porous medium can be written as:

$$S_s \frac{\partial h}{\partial t} = \frac{\partial}{\partial x}\left(K(x,y)\frac{\partial h}{\partial x}\right) + \frac{\partial}{\partial y}\left(K(x,y)\frac{\partial h}{\partial y}\right) \tag{1}$$

where $S_s$ is the specific storage; $h$ denotes the hydraulic head; and $K(x,y)$ means the hydraulic conductivity, which can be treated as a realization of a random field following a specific distribution with corresponding covariance. In this work, the Karhunen–Loeve expansion (KLE) is utilized to parameterize the heterogeneous model parameter with a determined covariance, as similarly done in TgNN (Wang et al., 2020). The TgGAN is, however, not limited to the KLE parameterization since $K(x,y)$ is regarded as an image therein, irrespective of how it is parameterized.

### 2.2 Generative Adversarial Network

Prior to introducing the main algorithms, we briefly review the architecture of GANs and WGAN-GP. GANs usually consist of a discriminator $D_\rho(\cdot)$ and a generator $G_\theta(\cdot)$, parameterized by $\rho$ and $\theta$, respectively. The GANs framework can be constructed based on ANN, CNN, and RNN. GANs can learn the distribution of target data based on the zero-sum game of $D_\rho(\cdot)$ and $G_\theta(\cdot)$. Taking the random latent vectors z-sampled from a specific distribution (e.g., Gaussian) as inputs, the $G_\theta(\cdot)$ can generate massive samples $G_\theta(z)$ that denote a new distribution $P_g$. The $D_\rho(\cdot)$ takes a new sample $x$ as input with an aim to determine whether the $D_\rho(x)$ is sampled from $P_g$ or a real data distribution $P_r$. Then, the $G_\theta(\cdot)$ will update its parameters to approximate $P_r$ with $P_g$ until the $D_\rho(\cdot)$ cannot distinguish $P_g$ and $P_r$. The structure of the standard GANs is illustrated in **Figure 1.**

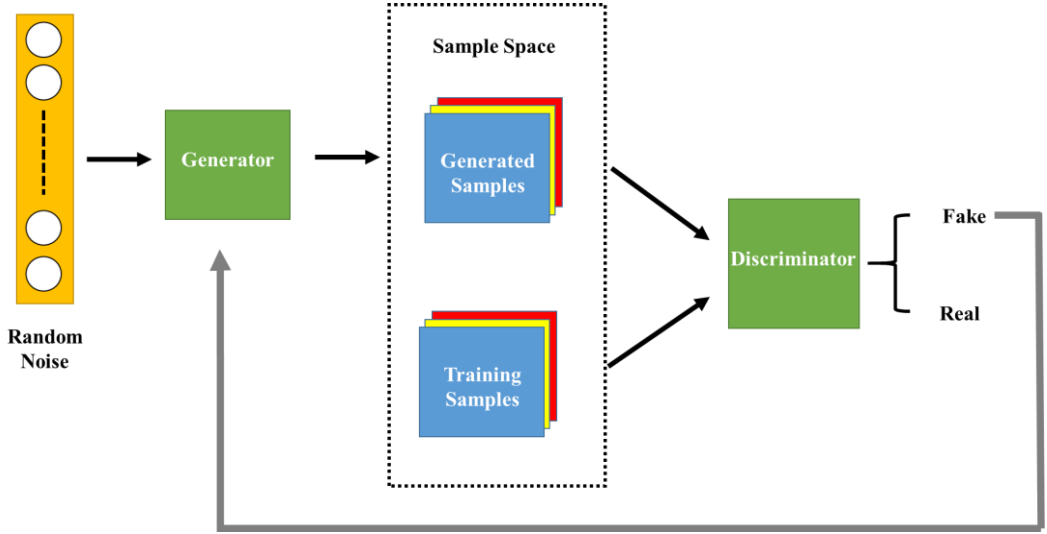

**Figure 1.** Structure of the GANs.

The loss functions of standard GANs are as follows (Goodfellow et al., 2014):

$$L_g = \mathbb{E}_{z \sim P_g} \left[ \log \left( 1 - D_\rho(G_\theta(z)) \right) \right]$$
$$L_d = -\mathbb{E}_{x \sim P_r} \left[ \log \left( D_\rho(x) \right) \right] - \mathbb{E}_{z \sim P_g} \left[ \log \left( 1 - D_\rho(G_\theta(z)) \right) \right]$$
(2)

where $x$ represents the samples from a real data distribution; $G_\theta(z)$ is generated samples; $\mathbb{E}_{x \sim P_r}$ and $\mathbb{E}_{z \sim P_g}$ stand for expectation.

If GANs are well trained, the loss function of generators can be expressed as Jensen-Shannon (JS) divergence (Goodfellow et al., 2014):

$$L_g = 2JS(P_r \parallel P_g) - 2\log 2 \tag{3}$$

where JS divergence can be expressed as Kullback-Leibler (KL) divergence (Goodfellow et al., 2014):

$$JS(P_r \parallel P_g) = \tfrac{1}{2} KL\left(P_r \parallel \tfrac{P_r + P_g}{2}\right) + \tfrac{1}{2} KL\left(P_g \parallel \tfrac{P_r + P_g}{2}\right) \tag{4}$$

In the training process, $P_g$ approximates $P_r$ by minimizing the loss function of the discriminator and the generator, respectively. The JS divergence, however, cannot provide effective gradients to update the generator when the overlap between $P_g$ and $P_r$ is difficult to capture. Consequently, mode collapse and training instability will occur. To solve these problems, weight-clipped Wasserstein GANs (WGANs) were proposed (Arjovsky et al., 2017).

The equation of Wasserstein distance, also called Earth Mover's distance, is as follows (Arjovsky et al., 2017):

$$W(P_r, P_g) = \inf_{\gamma \in \Gamma(P_r, P_g)} \mathbb{E}_{(P_r, P_g) \sim \gamma} [\lVert P_r - P_g \rVert] \tag{5}$$

where $\Gamma(P_r, P_g)$ is the set of all possible joint distributions of $P_r$ and $P_g$. Sampling from every possible joint distribution item $\gamma$, the distance between $P_r$ and $P_g$ and the expectation can be calculated, respectively. The lower bound that can be taken on this expectation in all possible joint distribution is Wasserstein distance.

In the training process, even if the supporting sets of the two distributions do not overlap or the overlap is very small, the Wasserstein distance can still reflect the distance between $P_g$ and $P_r$, to ensure that the gradients of the generator do not disappear, so as to solve the problem of mode collapse and training instability.

The loss functions of weight-clipped WGANs are defined as (Arjovsky et al., 2017):

$$L_g = -\mathbb{E}_{z \sim P_g}[D_\rho(G_\theta(z))]$$
$$L_d = -\mathbb{E}_{x \sim P_r}[D_\rho(x)] + \mathbb{E}_{z \sim P_g}[D_\rho(G_\theta(z))]$$
(6)

Clipping weight ensures the convergence of the loss functions. However, in the experimental stage, clipping weight can easily cause the parameters to be concentrated on both sides of the clipped position, resulting in uniformity of the generated results. To address this issue, Gulrajani et al. (2017) proposed WGAN with gradient penalty (WGAN-GP), which replaces the weight clipping method with adding a gradient penalty to the discriminator. Essentially, the gradient penalty term is the deformation of the K-Lipschitz continuity condition, which can be defined as follows:

$$L_p = \lambda \mathbb{E}_{\tilde{x} \sim P_{\tilde{x}}}\left[\left(\|\nabla D_\rho(\tilde{x})\|_2 - 1\right)^2\right]$$
(7)

where $P_{\tilde{x}}$ is the set of all possible joint distributions of $P_r$ and $P_g$; $\tilde{x}$ is sampled from $P_{\tilde{x}}$; and $\lambda$ is the coefficient of the gradient penalty term to weight different loss terms.

Finally, the loss functions of WGAN-GP are represented as (Gulrajani et al., 2017):

$$L_g = -\mathbb{E}_{z \sim P_g}[D_\rho(G_\theta(z))]$$
$$L_d = -\mathbb{E}_{x \sim P_r}[D_\rho(x)] + \mathbb{E}_{z \sim P_g}[D_\rho(G_\theta(z))] + \lambda \mathbb{E}_{\tilde{x} \sim P_{\tilde{x}}}\left[\left(\|\nabla D_\rho(\tilde{x})\|_2 - 1\right)^2\right]$$
(8)

Because of the stable training scheme, WGAN-GP is selected as the basic structure in this work.

**2.3 Theory-guided Generative Adversarial Network**

For standard GANs, therein referring to WGAN-GP, in which the inputs of generators are usually latent random vectors, there is no mapping between input and output. As a result, it is difficult for standard GANs to predict new responses, especially whose distributions vary with time.

In the dynamic subsurface flow problem, the distribution of hydraulic head changes spatiotemporally, which leads to the inability of GANs to learn the flow process, because the training data of GAN must satisfy a certain or similar distribution. In the proposed theory-guided generative adversarial network (TgGAN), the discriminator no longer judges the original data and the generated data, but rather their residuals. When the generated data are close to the training data, the residual matrix approximates a null matrix. In addition, underlying theories can be appended on the loss function of neural networks as soft constraints to improve accuracy and performance. In this study, the output of the generator is the hydraulic head, and the theory of flow in the porous medium in **Eq. (1)** can be rewritten as:

$$S_s \frac{\partial G_\theta(x,y,t)}{\partial t} = \frac{\partial}{\partial x}\left(K(x,y)\frac{\partial G_\theta(x,y,t)}{\partial x}\right) + \frac{\partial}{\partial y}\left(K(x,y)\frac{\partial G_\theta(x,y,t)}{\partial y}\right) \tag{9}$$

where $G_\theta(x, y, t)$ is the output of the generator. In TgGAN, CNN is chosen as the basic structure to build the generator and the discriminator.

The residual of the physical constraint can then be expressed as:

$$f \coloneqq S_s \frac{\partial G_\theta(x,y,t)}{\partial t} - \frac{\partial}{\partial x}\left(K(x,y)\frac{\partial G_\theta(x,y,t)}{\partial x}\right) - \frac{\partial}{\partial y}\left(K(x,y)\frac{\partial G_\theta(x,y,t)}{\partial y}\right) \tag{10}$$

where the derivatives of $G_\theta(\text{x}, \text{y}, \text{t})$ can be computed by numerical difference (e.g., Sobel filter) (Zhu et al., 2019) or automatic differentiation through networks. In this study, we employ the central difference method to calculate the derivatives. Some collocation points, rather than the entire field, are chosen to construct the physical constraints.

The mean square error (MSE) of the physical constraint is added to the loss function of the generator as an additional part:

$$L_{\text{PDE}} = \frac{1}{N}\sum_{i=1}^{N}\left|S_s \frac{\partial G_\theta(x_i,y_i,t_i)}{\partial t_i} - \frac{\partial}{\partial x_i}\left(K(x_i,y_i)\frac{\partial G_\theta(x_i,y_i,t_i)}{\partial x_i}\right) - \frac{\partial}{\partial y_i}\left(K(x_i,y_i)\frac{\partial G_\theta(x_i,y_i,t_i)}{\partial y_i}\right)\right|^2 \tag{11}$$

where $N$ is the number of collocation points; and $x_i$, $y_i$, and $t_i$ mean the x-coordinate, y-coordinate, and time step of the $i_{th}$ collocation point, respectively. We combine *x*, *y*, and *t* into an image with three channels at each time step to generate the corresponding h with one channel. Other theories, such as engineering controls and expert knowledge, can also be encoded into the loss function of the generator:

$$L_{\text{EC}} = \frac{1}{N_{\text{EC}}}\sum_{i=1}^{N_{\text{EC}}}\left|f_{\text{EC}}(x^i, y^i, t^i)\right|^2 \tag{12}$$

$$L_{\text{EK}} = \frac{1}{N_{\text{EK}}}\sum_{i=1}^{N_{\text{EK}}}\left|f_{\text{EK}}(x^i, y^i, t^i)\right|^2 \tag{13}$$

where $L_{\text{EK}}$ is the loss term of expert knowledge; $L_{\text{EC}}$ denotes the loss term of engineering controls; and $f_{\text{EK}}$ and $f_{\text{EC}}$ is the function of engineering control and expert knowledge, respectively.

The residual of boundary conditions (e.g., inlet condition, outlet condition and no flow boundary) and initial conditions are expressed as follows:

$$f_{\text{BC}} \coloneqq G_\theta(x_{\text{BC}}, y_{\text{BC}}, t) - h_{\text{BC}} \tag{14}$$

$$f_{\text{IC}} \coloneqq G_\theta(x, y, t_{\text{IC}}) - h_{\text{IC}} \tag{15}$$

The MSE of boundary and initial conditions are also integrated in the form of soft constraints, as shown below:

$$L_{\text{BC}} = \frac{1}{N_{\text{BC}}}\sum_{i=1}^{N_{\text{BC}}}\left|f_{\text{BC}}(x^i_{\text{BC}}, y^i_{\text{BC}}, t^i)\right|^2 \tag{16}$$

$$L_{\text{IC}} = \frac{1}{N_{\text{IC}}}\sum_{i=1}^{N_{\text{IC}}}\left|f_{\text{IC}}(x^i, y^i, t^i_{\text{IC}})\right|^2 \tag{17}$$

Thus, the loss functions of TgGAN can be expressed as:

$$L_g = -\mathbb{E}_{z\sim P_g}\left[D_\rho\left(R_g(t_j)\right)\right] + \lambda_{\text{PDE}}L_{\text{PDE}} + \lambda_{\text{BC}}L_{\text{BC}} + \lambda_{\text{IC}}L_{\text{IC}} + \lambda_{\text{EK}}L_{\text{EK}} + \lambda_{\text{EC}}L_{\text{EC}}$$
$$L_d = -\mathbb{E}_{x\sim P_r}\left[D_\rho\left(R_r(t_j)\right)\right] + \mathbb{E}_{z\sim P_g}\left[D_\rho\left(R_g(t_j)\right)\right] + \lambda\mathbb{E}_{\tilde{x}\sim P_{\tilde{x}}}\left[\left(\|\nabla D_\rho(\tilde{x})\|_2 - 1\right)^2\right]$$
(18)

where the $R_g(t_j) = G_\theta(t_j) - h(t_j)$ is the residual matrix between the training data and the generated one at the time step $j$ ( $j=0....k$, $k$ is the maximum time step of training data sets); $R_r(t_j)$ is a null matrix; and $\mathbb{E}_{x\sim P_r}$ and $\mathbb{E}_{x\sim P_g}$ stand for expectation. In the training process, the generator will generate $n$ predictions ($n \geq k$). $n$ is the maximum time step of the whole flow process, but the residual matrixes, only between the first $k$ generated data and the real data, are inputted into the discriminator and compared to the null matrix. However, the collocation points can be taken in whole n time steps, which means that the number of data points and collocation points can be different. Therefore, the predictions of TgGAN of the first $k$ time steps can be regarded as interpolation, while the predictions of the last $n$-$k$ time steps are extrapolation. $\lambda_{\text{PDE}}$, $\lambda_{\text{IC}}$, $\lambda_{\text{BC}}$, $\lambda_{\text{EK}}$, and $\lambda_{\text{EC}}$ are the hyper-parameters to determine the weight of each term. The structure of the TgGANs is illustrated in **Figure 2.**

It is the case that other neural networks, such as ANN and RNN, may be used. In this work, however, the discriminator and generator are both implemented with CNN, which is a kind of neural network that can process image data efficiently. Although the structure of discriminator is slightly different for different amounts of training data, the generator is the same because we aim to use it to generate all of the data at once. **Subsection 2.4** briefly introduces basic knowledge about CNN.

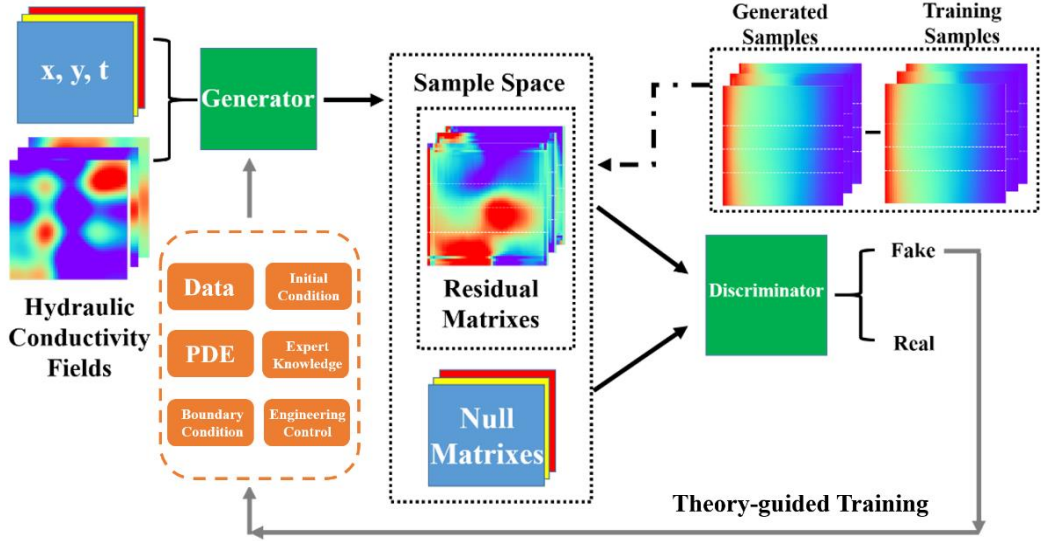

**Figure 2.** Structure of the TgGAN.

## 2.4 Convolutional Neural Network

The properties of an image include the number of channels and pixels. Even though artificial neural networks (ANNs) have achieved notable success in classification and regression problems by connecting neural units with each other, when the input layer is characterized by high-dimensional images, the parameters of ANN will become exceedingly large, and the training process will be computationally demanding. The convolutional neural network (CNN) is an efficient network for processing image data. CNN uses the convolutional operation to share connection weights $w$ and biases $b$ between neurons, which greatly decreases the number of training parameters to reduce the dimension. CNN usually includes one input layer, several convolutional layers, activation layers, pooling layers, fully connected layers, and an output layer. In the convolutional calculation process, a padding operation can be used to expand or maintain the dimension, and $p$-loop 0's are filled around the data outputted from the previous step. In the convolutional layers, kernels traverse the image according to the stride $s$. The two-dimensional convolutional calculation process of an image with two channels is shown in **Figure 3**, where $i_{11}^1 \dots i_{55}^2$ denote original pixels of the input image; $k_{11}^1 \dots k_{33}^3$ stand for pixels of the kernel; $o_{11}^1 \dots o_{22}^3$ mean the pixels of the output image; and * means the convolutional operation. The relationship between the size of input image pixels and the size of output image pixels is shown as follows:

$$o = \frac{i-k+2p}{s} + 1 \tag{19}$$

where $i$ is the size of the input image; $k$ is the kernel size; $p$ is the one-side width of padding; $s$ is the stride of the convolutional kernel moving over the original image; and $o$ denotes the size of the output image. In **Figure 3**, $i$ is 5, $k$ is 3, $p$ is 0, and $s$ is 2. Thus, according to **Eq. (19)**, $o$ is 2.

The convolutional process in **Figure 3** is:

$$\begin{aligned}
o_{11}^1 &= i_{11}^1 k_{11}^1 + i_{12}^1 k_{12}^1 + i_{13}^1 k_{13}^1 + i_{21}^1 k_{21}^1 + i_{22}^1 k_{22}^1 + i_{23}^1 k_{23}^1 + i_{31}^1 k_{31}^1 + i_{32}^1 k_{32}^1 \\
&\quad + i_{33}^1 k_{33}^1 \\
o_{12}^1 &= i_{13}^1 k_{11}^1 + i_{14}^1 k_{12}^1 + i_{15}^1 k_{13}^1 + i_{23}^1 k_{21}^1 + i_{24}^1 k_{22}^1 + i_{24}^1 k_{23}^1 + i_{33}^1 k_{31}^1 + i_{34}^1 k_{32}^1 \\
&\quad + i_{35}^1 k_{33}^1 \\
o_{21}^1 &= i_{31}^1 k_{11}^1 + i_{32}^1 k_{12}^1 + i_{33}^1 k_{13}^1 + i_{41}^1 k_{21}^1 + i_{42}^1 k_{22}^1 + i_{43}^1 k_{23}^1 + i_{51}^1 k_{31}^1 + i_{52}^1 k_{32}^1 \\
&\quad + i_{53}^1 k_{33}^1 \\
o_{22}^1 &= i_{33}^1 k_{11}^1 + i_{34}^1 k_{12}^1 + i_{35}^1 k_{13}^1 + i_{43}^1 k_{21}^1 + i_{44}^1 k_{22}^1 + i_{45}^1 k_{23}^1 + i_{53}^1 k_{31}^1 + i_{54}^1 k_{32}^1 \\
&\quad + i_{55}^1 k_{33}^1
\end{aligned} \tag{20}$$

$o_{11}^2, o_{12}^2 \dots o_{21}^3$ and $o_{22}^3$ are calculated in the same manner. The number of channels for the output image is determined by the number of kernels. For example, if the number of kernels is $n$, the number of channels of the output image will be $n$ correspondingly.

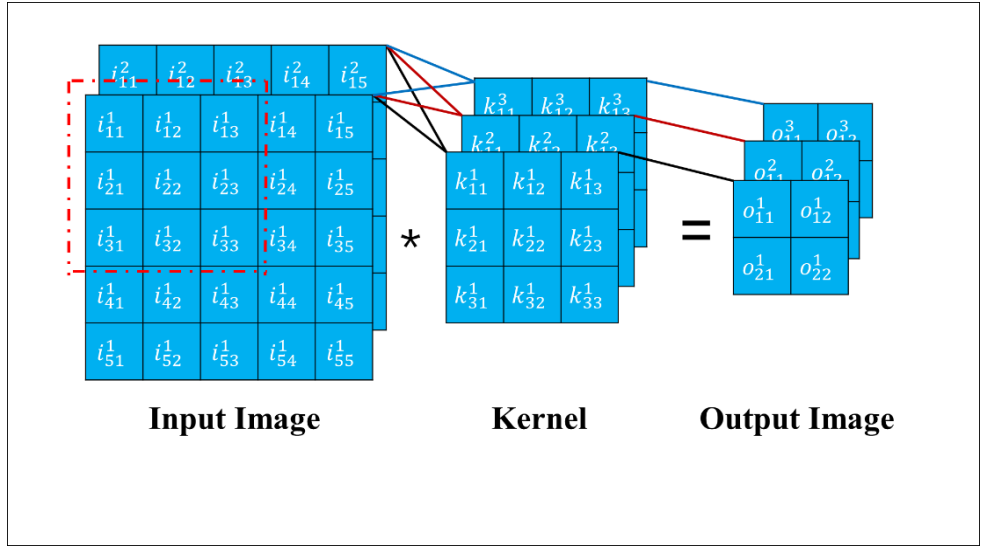

**Figure 3.** Convolutional calculation process in CNN

In the TgGAN model, input variables *x, y,* and *t* are combined as an image with three channels if the hydraulic conductivity field *K* is given. For the case of predicting corresponding hydraulic heads for new *K* fields, the number of input channels will change to four because the *K* field should be combined with coordinates to provide additional information. The hydraulic head *h* is regarded as an image with one channel at each time step. In the training process, the input dimension of the generator is $T \times C \times R \times L$, where $T$ is the number of time steps; $C$ is the number of channels; $R$ is the number of rows of the input image; and $L$ is the number of columns.

The discriminator for each case has four layers, while the size of the kernel, stride, and padding vary slightly with the number of training data. Meanwhile, the setup for the generator is fixed.

An overview of the structures of the discriminator and the generator are shown in **Table 1** and **Table 2**, respectively.

**Table 1.** Structures of the discriminator.

| Layer | Channel (c) | Kernel (*k*) | Stride (*s*) | Padding (*p*) |
|---|---|---|---|---|
| 1 | 1 | $k_1$ | $s_1$ | $p_1$ |
| 2 | 16 | $k_2$ | $s_2$ | $p_2$ |
| 3 | 16 | $k_3$ | $s_3$ | $p_3$ |
| 4 | 1 | $k_4$ | $s_4$ | $p_4$ |

where $k_i, S_i,$ and $P_i$ (*i*=1, 2, 3, 4) is the kernel size, stride, and padding for different cases, respectively.

Table 2. Structures of the generator.

| Layer | Channel (c) | Kernel (k) | Stride (s) | Padding (p) |
|---|---|---|---|---|
| 1 | 3 | 3 | 1 | 1 |
| 2 | 32 | 3 | 1 | 1 |
| 3 | 32 | 3 | 1 | 1 |
| 4 | 32 | 3 | 1 | 1 |
| 5 | 32 | 3 | 1 | 1 |
| 6 | 1 | 3 | 1 | 1 |

## 3. Case Studies

In this section, the performance of TgGAN is tested by several subsurface flow cases. The accuracy of TgGAN is also compared with standard GAN without physical constraints.

The domain is a square divided into 51×51 grids with 1020 [L] in both directions. The left side is the entrance and the right side is the exit, taking values of $H_{x=0} = 5[L]$ and $H_{x=1020} = 0[L]$, respectively, unless otherwise stated. The upper and lower boundaries are impervious. $S_s$ is set as a constant of 0.0001 [$L^{-1}$]. The total flow duration is set to 8 [T] and the time step to 0.1 [T], so the generator will produce 80 hydraulic head images, unless otherwise stated. The training database is generated by simulation software MODFLOW.

The activation function for each layer is LeakyReLU for both the generator and the discriminator, and the optimizer is Adam (Kingma et al., 2015) with a learning rate of 0.0001. In **subsections 3.1** to **3.5**, the input dimensions of the generator are all 80×3×51×51 because the generator will generate *h* for the whole flow process, and then data points and collocation points will be chosen, respectively. The input dimensions of the generator are different in **subsections 3.6** and **3.7**. Data points will be treated as the input of the discriminator. The entire program is run in the Pytorch environment, and the training process is carried out on an NVIDIA GeForce RTX 2080 Ti.

### 3.1 Predicting the future response

In this case, the hydraulic head distribution at the forehand 20 time steps is monitored, and 625 (25×25) points are chosen randomly as training data for each of the first 20 time steps. In the entire 80 time steps, 576 (24×24) points are extracted as collocation points at every other time step, amounting to a total of 23,040 (i.e., 576×40 time steps) collocation points. Both the collocation points and the data points are randomly selected in space, but the positions of these points are fixed with respect to time. The distributions at the first 20 time steps are to be interpolated and those at the following time steps are expected to be extrapolated.

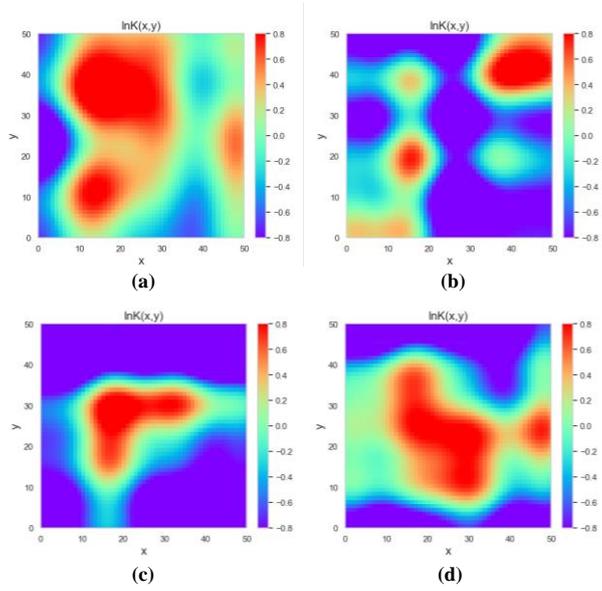

**Figure 4.** Four hydraulic conductivity fields (a, b, c, and d).

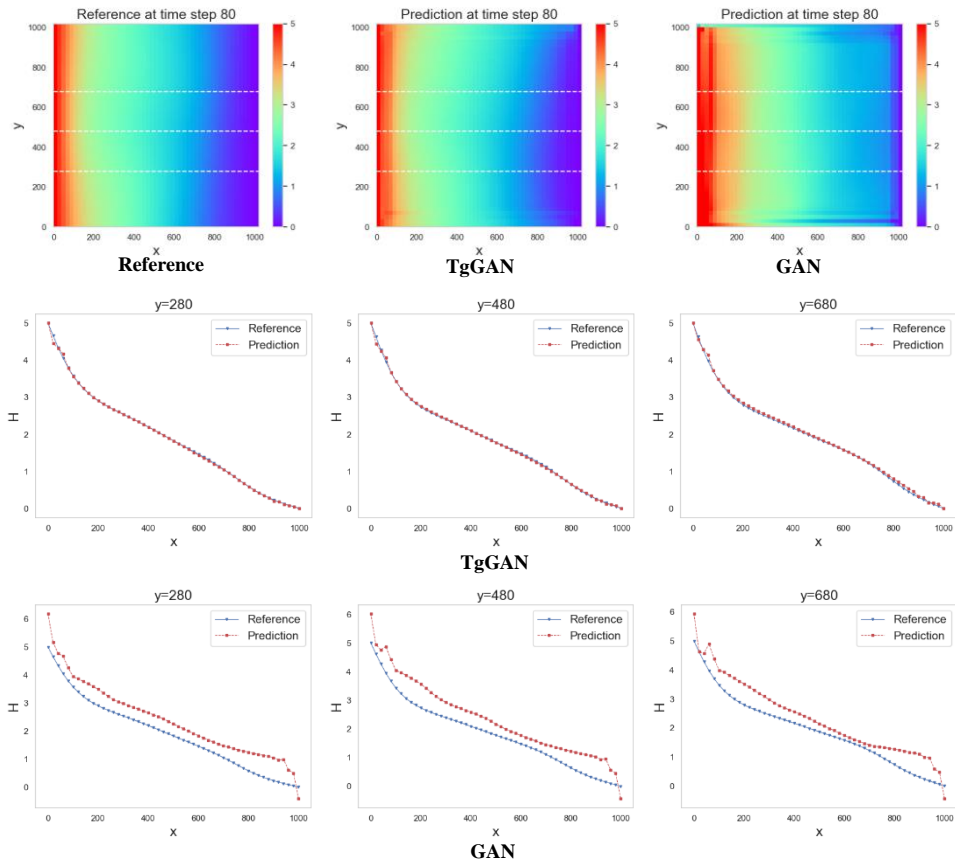

**Figure 5.** Predictions of TgGAN and GAN for hydraulic conductivity field (**Figure 4a**).

We use four cases with different hydraulic conductivity (*ln K*) fields (a, b, c, and d), as shown in **Figure 4**, to test the performance of TgGAN. The same training data from the first 20 time steps are also used for training the standard GAN, which is not required to honor the equation, as well as the boundary and initial conditions (thus, no collocation points are employed). The predictions for the hydraulic conductivity field (**Figure 4a**) given by TgGAN and GAN at time step 80 are presented in **Figure 5**, and the predictions for other fields (**Figure 4b-4d**) are shown in **Appendix A.1**. **Figure 6** presents the scatterplots of the hydraulic head predicted by TgGAN and GAN at time step 80. It can be shown that the predictions of TgGAN are much more accurate than those of GAN.

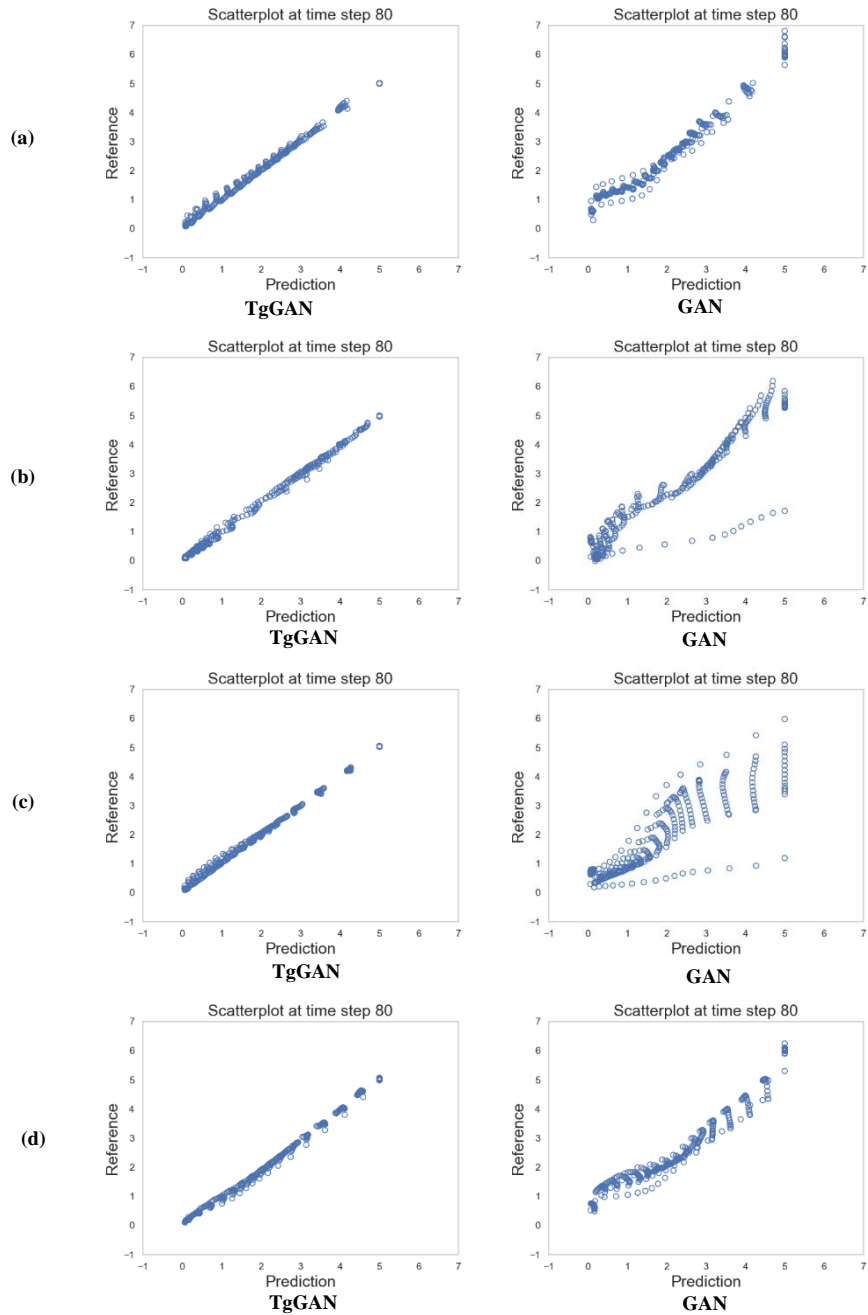

**Figure 6.** Scatterplots of the hydraulic head predicted by TgGAN and GAN at time step 80 for four hydraulic conductivity fields (a, b, c, and d).

Furthermore, relative $L_2$ error and $R^2$ score are used to evaluate the results of TgGAN and GAN, which are defined as follows:

$$L_2(u_{pred}, u_{real}) = \frac{\|u_{pred} - u_{real}\|_2}{\|u_{real}\|_2} \tag{21}$$

$$R^2 = 1 - \frac{\sum_{n=1}^{N}(u_{pred,n} - u_{real,n})^2}{\sum_{n=1}^{N}(u_{real,n} - \bar{u}_{real})^2} \tag{22}$$

where $u_{pred}$ and $u_{real}$ are the prediction and the reference, respectively; and $\bar{u}_{real}$ denotes the mean of $u_{real}$. The means and variances of relative $L_2$ error and $R^2$ score of the whole 80 time steps of TgGAN and GAN for four hydraulic conductivity fields (a, b, c, and d) are shown in **Table 3**. It is seen that TgGAN outperforms GAN significantly with reduced relative $L_2$ error and increased $R^2$ score. For TgGAN the training time is longer than GAN, from 9.68% to 22.52%, due to the physical constraints employed.

**Table 3.** Means and variances of relative $L_2$ error and $R^2$ score of the whole 80 time steps of TgGAN and GAN for four hydraulic conductivity fields (a, b, c, and d).

|  | TgGAN | | | | |
| --- | --- | --- | --- | --- | --- |
|  | Relative $L_2$ error | | $R^2$ score | | **Running time** |
|  | **Mean** | **Variance** | **Mean** | **Variance** |  |
| **Field (a)** | 2.7540e-2 | 2.1000e-4 | 0.9966 | 7.2811e-6 | 2,040s |
| **Field (b)** | 2.9778e-2 | 2.0100e-4 | 0.9970 | 4.4010e-6 | 1,869s |
| **Field (c)** | 2.5574e-2 | 1.2100e-4 | 0.9978 | 2.3502e-6 | 1,787s |
| **Field (d)** | 4.4515e-2 | 9.6400e-5 | 0.9942 | 8.9583e-6 | 1,949s |
|  | GAN | | | | |
| **Field (a)** | 1.1907e-1 | 1.5200e-3 | 0.9481 | 1.8253e-3 | 1,665s |
| **Field (b)** | 1.4148e-1 | 5.1000e-4 | 0.9455 | 9.3011e-4 | 1,704s |
| **Field (c)** | 1.4925e-1 | 1.9180e-3 | 0.9165 | 5.2387e-3 | 1,601s |
| **Field (d)** | 1.0342e-1 | 1.3120e-3 | 0.9560 | 1.7608e-3 | 1,664s |

**Figure 7** presents the curves of relative $L_2$ error and $R^2$ score over time for TgGAN and GAN for four hydraulic conductivity fields. The first 20 time steps are for interpolation, while the subsequent 60 time steps are for extrapolation. It can be seen that the relative $L_2$ errors of TgGAN for four hydraulic conductivity fields are smaller than those of GAN. Furthermore, the $R^2$ scores of TgGAN are closer to 1 than those of GAN, indicating that the TgGAN model can match the reference better. It is also seen that the performance of TgGAN is more stable because it can achieve similar accuracy in both of the interpolation and the extrapolation stages, while the standard GAN has obviously inferior results in the extrapolation stage than that in the interpolation stage. The underlying reason for this is that the physical constraints imposed in TgGAN enhance its robustness for temporal extrapolation. At the first several time steps, the randomly selected training data and collocation points may not be representative for the dynamic flow since the particular setup is confined to a small area near the left boundary, which explains the relatively poor quality at the early times for both GAN and TgGAN.

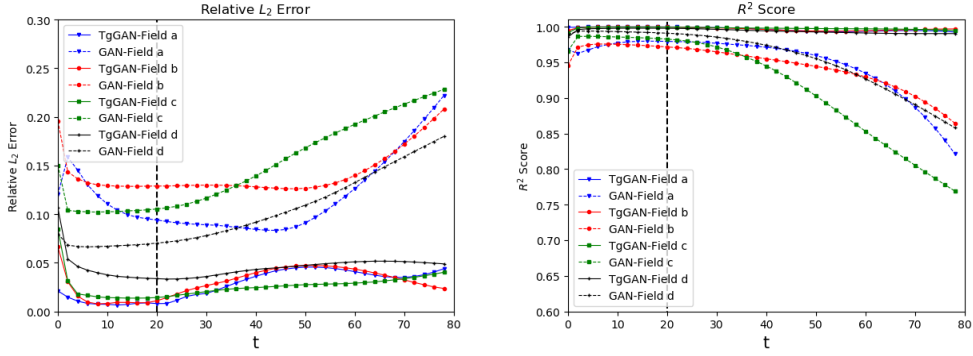

**Figure 7.** Change of relative $L_2$ error and $R^2$ score over time for TgGAN and GAN for four hydraulic conductivity fields (a, b, c, and d).

### 3.2 Predicting the future response from noisy data

The training data in **subsection 3.1** are clean or noiseless. However, most of the time, data are accompanied by more or less noise, which will affect the accuracy of prediction. In this subsection, we test the ability of TgGAN and GAN to resist different levels of noise, defined as:

$$\mathrm{h}(x, y, t) = \mathrm{h}(x, y, t) + h_{diff}(x, y) \times a\% \times \varepsilon \qquad (19)$$

where $h_{diff}(x, y)$ is the maximum difference of training data; $a$ is the percentage of noise; and $\varepsilon$ denotes a random value ranging from -1 to 1.

Field (c) in **subsection 3.1** is chosen to train TgGAN and GAN. **Figure 8** shows the profiles of the hydraulic head with 5%, 10% and 20% noise, and the scatterplots of the hydraulic head for TgGAN and GAN at time step 80, respectively. The predictions at time step 80 of TgGAN and GAN trained with noisy data, as well as corresponding distributions, are shown in **Appendix A.2**. It is obvious that the results of the GAN model are seriously affected by noise, while TgGAN is more stable and reliable owing to the enforcement of physical constraints. The relative $L_2$ errors and $R^2$ scores over time for TgGAN and GAN with different levels of noise are presented in **Figure 9**. The means and variances of relative $L_2$ error and $R^2$ score of the whole 80 time steps of TgGAN and GAN when noise exists are shown in **Table 4**.

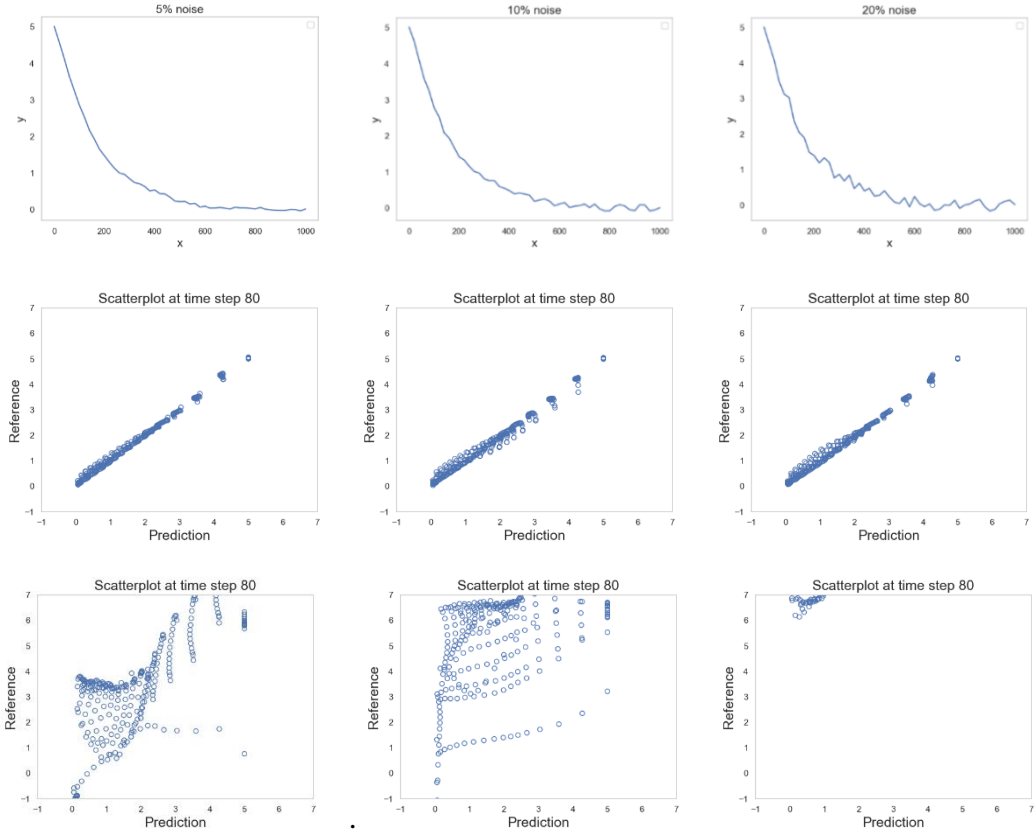

**Figure 8.** Profiles of the hydraulic head with 5%, 10% and 20% noise, and the scatterplots of hydraulic head for TgGAN and GAN at time step 80. First row: profiles of the hydraulic head with 5%, 10% and 20% noise. Second row: scatterplots of the hydraulic head for TgGAN. Third row: scatterplots of the hydraulic head for GAN.

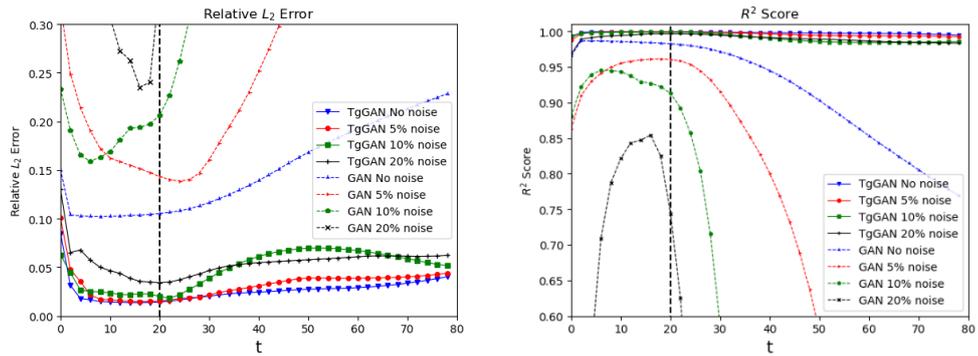

**Figure 9.** Changes of relative $L_2$ error and $R^2$ score over time for TgGAN and GAN when noise exists.

**Table 4.** Means and variances of relative $L_2$ error and $R^2$ score of the whole 80 time steps of TgGAN and GAN when noise exists.

|  | TgGAN | | | | Running time |
|---|---|---|---|---|---|
|  | Relative $L_2$ error | | $R^2$ score | | |
|  | Mean | Variance | Mean | Variance | |
| **No noise** | 2.5574e-2 | 1.2100e-4 | 0.99783 | 2.3502e-6 | 1,787s |
| **5% noise** | 3.1649e-2 | 1.7800e-4 | 0.99612 | 6.7241e-6 | 1,784s |
| **10% noise** | 4.8070e-2 | 3.4100e-4 | 0.99131 | 3.7119e-5 | 1,896s |
| **20% noise** | 5.4244e-2 | 1.6400e-4 | 0.99008 | 2.5330e-5 | 1,898s |
|  | GAN | | | | |
| **No noise** | 1.4925e-1 | 1.9180e-3 | 0.91646 | 5.2387e-3 | 1,601s |
| **5% noise** | 3.6510e-1 | 5.4510e-2 | 0.47348 | 4.1205e-1 | 1,580s |
| **10% noise** | 8.6022e-1 | 3.3652e-1 | -1.6508 | 7.8707 | 1,662s |
| **20% noise** | 1.9432 | 2.0018 | -13.9092 | 304.2069 | 1,743s |

It is obvious that TgGAN is more resistant to noise compared to GAN. It is difficult for GAN to learn the correct distribution from noisy data; however, with the assistance of physical constraints, this problem is alleviated. It is worth noting to note that the training time for TgGAN is only slightly increased compared to GAN for this particular hydraulic conductivity field.

### 3.3 Predicting the future response without labels

In this subsection, we aim to train TgGAN to generate the whole flow process without any labels. The hydraulic conductivity field (d) in **subsection 3.1** is chosen as the corresponding field.

For the discriminator, the input is only the residual between the generated initial condition and the real one. The number of collocation points at every other step during the whole process of 80 time steps is 625, amounting to a total of 25,000 (i.e., 625×40 time steps) collocation points. The results of TgGAN at time step 80 are shown in **Figure 10**. The change of relative $L_2$ error and $R^2$ score over time for TgGAN without labels are presented in **Figure 11**. The means and variances of relative $L_2$ error and $R^2$ score of the whole 80 time steps are shown in **Table 5**. The predictions of the hydraulic head are shown in **Appendix A.3**.

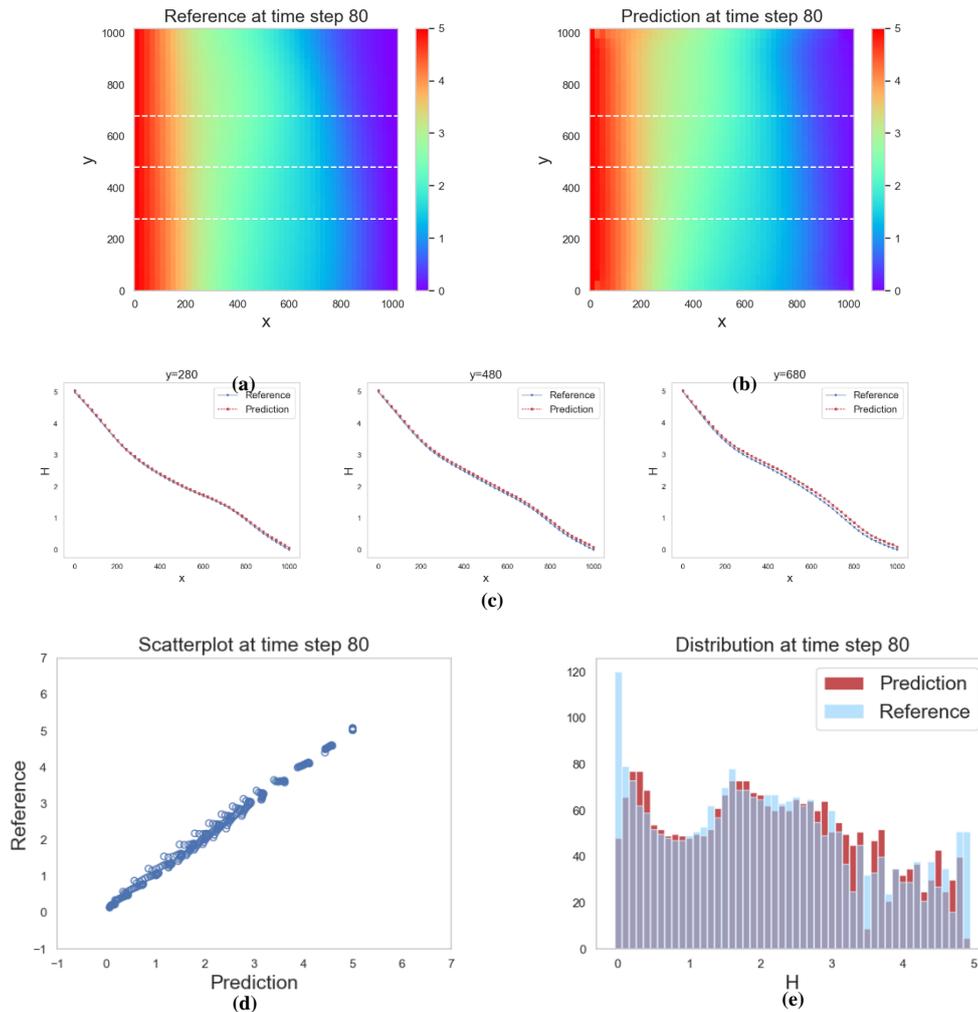

**Figure 10.** Results at time step 80 without labels. (a): reference hydraulic head at time step 80; (b): predicted hydraulic head at time step 80; (c): comparison of real data and predictions along three lines; (d): scatterplots; (e): distribution of hydraulic head.

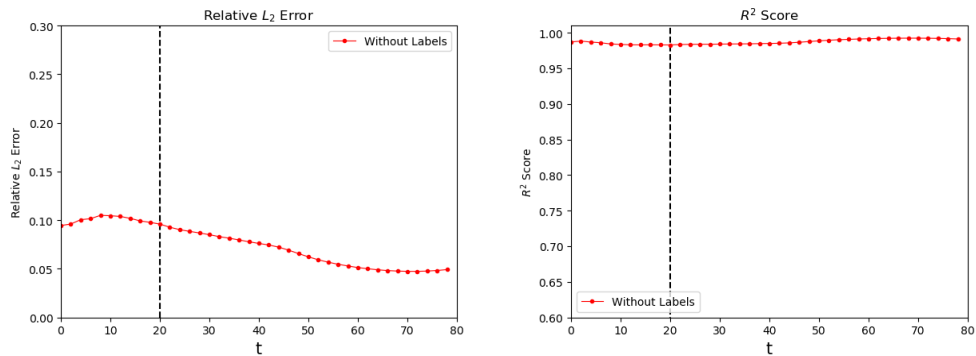

**Figure 11.** Changes of relative $L_2$ error and $R^2$ score over time for TgGAN without labels.

**Table 5.** Means and variances of relative $L_2$ error and $R^2$ score of the whole 80 time steps of TgGAN without labels.

| Relative $L_2$ error | | $R^2$ score | | Running time |
|---|---|---|---|---|
| **Mean** | **Variance** | **Mean** | **Variance** | 2,747s |
| 7.4558e-2 | 4.2500e-4 | 0.98726 | 1.2427e-5 | |

It can be seen that, in the absence of any training data, TgGAN can still generate the entire flow process with good accuracy via label-free learning by adhering to the physical constraints. In other words, TgGAN can be used for solving PDEs. However, compared to traditional numerical methods, computational efficiency is subject to investigation. Furthermore, training time is significantly longer without training data than the counterpart shown in **Table 3**. Essentially, label-free learning is possible, but comes with a larger computational demand, compared to that with labeled data. Therefore, we may recommend to take full use of training data to improve efficiency when quality data are available.

**3.4 Effect of the number of data points**

In the first two subsections, the number of training data points is 625 (25×25) at each of the first 20 time steps. Although a small number of data points and collocation points have achieved a high rate of accuracy, we would like to investigate whether the performance of TgGAN can be further improved by increasing the number of data points and collocation points. In this subsection, the effect of the number of training data on the results is explored. Moreover, the following **subsection 3.5** will discuss the effect of the number of collocation points. In this subsection and **subsection 3.5**, the hydraulic conductivity field (c) in **subsection 3.1** is chosen as the corresponding field.

We first control the number of collocation points to be 25,000 (i.e., 625×40 time steps), and increase the number of training data points at each of the first 20 time steps to 1,225 (35×35) and 2,601 (51×51) to test the performance of TgGAN, respectively.

The changes of relative $L_2$ error and $R^2$ score over time for TgGAN and GAN are shown in **Figure 12**. The means and variances of relative $L_2$ error and $R^2$ score of the whole 80 time steps for TgGAN and GAN with different numbers of data points are presented in **Table 6**.

It can be seen that the performance of both TgGAN and GAN improves during interpolation with increasing the number of training data. Such an improvement with data cannot be sustained for extrapolation over time for GAN. The quality of extrapolation for TgGAN does not improve either for this case, perhaps because the number of collocation points is inadequate. This issue will be revisited in the next subsection. It is worth noting that increasing the number of data points has little effect on the training time, because the hydraulic head is regarded as an image to be processed by CNN in both TgGAN and GAN.

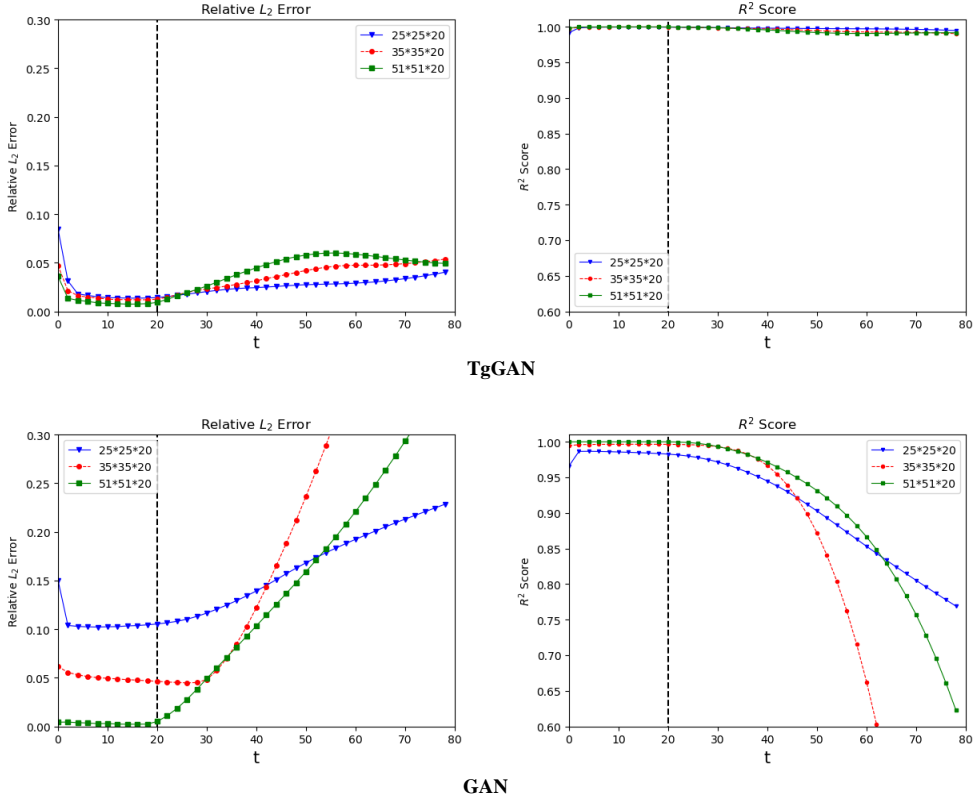

**Figure 12.** Changes of relative $L_2$ error and $R^2$ score over time for TgGAN and GAN with different numbers of data points.

**Table 6.** Means and variances of relative $L_2$ error and $R^2$ score of the whole 80 time steps for TgGAN and GAN with different numbers of data points.

|  | TgGAN | | | | |
| --- | --- | --- | --- | --- | --- |
| **Number of data points** | **Relative $L_2$ error** | | **$R^2$ score** | | |
|  | **Mean** | **Variance** | **Mean** | **Variance** | **Running time** |
| **15×15×20** | 2.5574e-2 | 1.2100e-4 | 0.99783 | 2.3502e-6 | 1,764s |
| **35×35×20** | 3.2210e-2 | 2.1600e-4 | 0.99613 | 1.0265e-5 | 1,787s |
| **51×51×20** | 3.6345e-2 | 3.9931e-4 | 0.99543 | 1.4521e-5 | 1,962s |
|  | GAN | | | | |
| **15×15×20** | 1.4925e-1 | 1.9180e-3 | 0.91646 | 5.2387e-3 | 1,561s |
| **35×35×20** | 2.1445e-1 | 3.8261e-2 | 0.78061 | 1.0841e-1 | 1,601s |
| **51×51×20** | 1.2482e-1 | 1.3355e-2 | 0.91851 | 1.1701e-2 | 1,664s |

### 3.5 Effect of the number of collocation points

In this subsection, we control the number of data points to be 625, 1,225, or 2,601 at each of the first 20 time steps and change the number of collocation points at every other step of the 80 time steps to 225 (15×15), 625 (25×25), and 1,600 (40×40), respectively, to explore the effect of the number of collocation points on the accuracy of hydraulic head predictions.

The changes of relative $L_2$ error and $R^2$ score over time for TgGAN with different numbers of collocation points for the case of 625, 1,225, and 2,601 data points at each of the first 20 time steps are presented in **Figure 13**. The means and variances of relative $L_2$ error and $R^2$ score of the whole 80 time steps for TgGAN with different numbers of collocation points for the case of 625, 1,225, and 2,601 data points at each of the first 20 time steps are shown in **Table 7**. It can be seen that the more are the collocation points, the better is the performance of TgGAN.

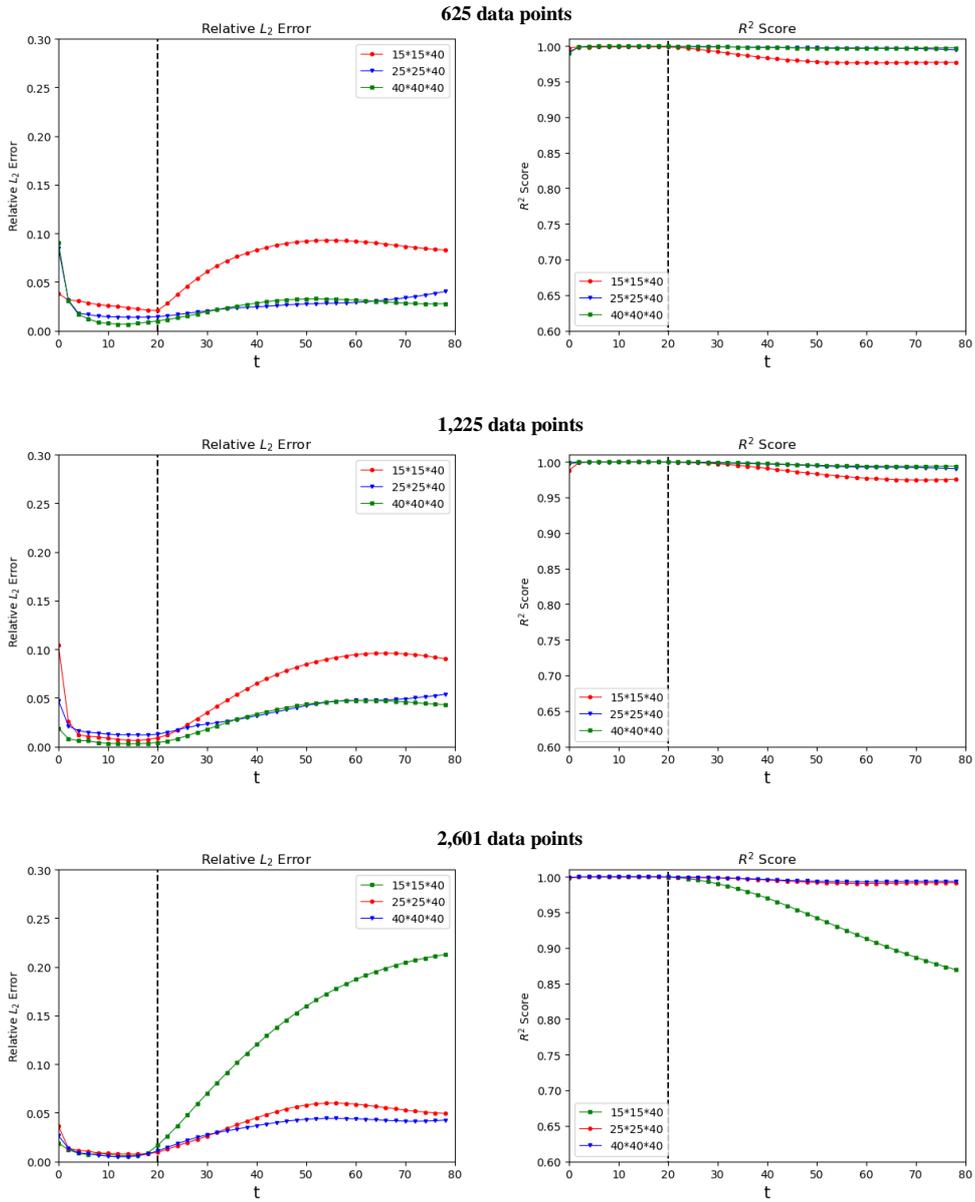

**Figure 13.** Changes of relative $L_2$ error and $R^2$ score over time for TgGAN with different numbers of collocation points for the case of 625, 1,225 and 2,601 data points at each of the first 20 time steps.

**Table 7.** Means and variances of relative $L_2$ error and $R^2$ score of the whole 80 time steps for TgGAN with different numbers of collocation points for the case of 625, 1,225 and 2,601 data points at each of the first 20 time steps.

| | TgGAN | | | | |
|---|---|---|---|---|---|
| | 625 data points | | | | |
| **Number of collocation points** | **Relative $L_2$ error** | | **$R^2$ score** | | **Running time** |
| | **Mean** | **Variance** | **Mean** | **Variance** | |
| **15×15×40** | 6.476e-2 | 7.7800e-4 | 0.98653 | 9.0563e-5 | 1,732s |
| **25×25×40** | 2.5574e-2 | 1.2100e-4 | 0.99783 | 2.3502e-6 | 1,746s |
| **40×40×40** | 2.3925e-2 | 1.4400e-4 | 0.99801 | 2.2825e-6 | 1,790s |
| | **1,225** data points | | | | |
| **15×15×40** | 5.6557e-2 | 1.2650e-3 | 0.98871 | 1.0257e-4 | 1,777s |
| **25×25×40** | 3.2101e-2 | 2.1600e-4 | 0.99613 | 1.0265e-5 | 1,787s |
| **40×40×40** | 2.7463e-2 | 3.1400e-4 | 0.99714 | 6.6101e-6 | 1,946s |
| | **2,601** data points | | | | |
| **15×15×40** | 1.0840e-1 | 6.0560e-3 | 0.95524 | 2.0603e-3 | 1,957s |
| **25×25×40** | 3.6345e-2 | 3.9993e-4 | 0.99543 | 1.4521e-5 | 1,962s |
| **40×40×40** | 2.9419e-2 | 2.1846e-4 | 0.99639 | 8.1290e-6 | 1,959s |

It is seen that, in general, the quality of prediction improves during both interpolation and extrapolation with increasing the number of collocation points for all cases of different training data. As shown in **Table 7**, when the number of training data is large (e.g., 2,601 data), the performance of interpolation is largely independent of the number of collocation points. The quality of extrapolation, however, can be worse for a large number of training data if the number of collocation points is not adequate. This is the case because the learning process gains false confidence from training data (at early times) without strictly satisfying the physical constraints for the entire process. This suggests that the number of collocation points should be commensurate with the number of training data. Again, it is worth noting that, in general, the training time increases slightly with both the number of data points and the number of collocation points.

### 3.6 Predicting the corresponding hydraulic heads for new hydraulic conductivity fields

In this subsection, the TgGAN that is trained with a specific hydraulic conductivity field is used for predicting the dynamic hydraulic head fields for new conductivity fields. The hydraulic conductivity field (a) in **subsection 3.1** is selected for performing this case study. The input dimension of the generator is $80 \times 4 \times 51 \times 51$. There are four input channels denoting four features, i.e., *t, K, x*, and *y*. We randomly chose 625 points as training data for each of the first 20 time steps, amounting to a total of 12,500 data points (i.e., 625×20 time steps). 2,209 points for each of the whole 80 time steps are extracted as the collocation points, amounting to a total of 176,720 (i.e., 2,209×80 time steps) collocation points. **Figure 14** shows the scatterplots of the hydraulic head predicted by TgGAN for four hydraulic conductivity fields (a, b, c, and d). The changes of relative $L_2$ error and $R^2$ score over time for TgGAN are presented in **Figure 15.**

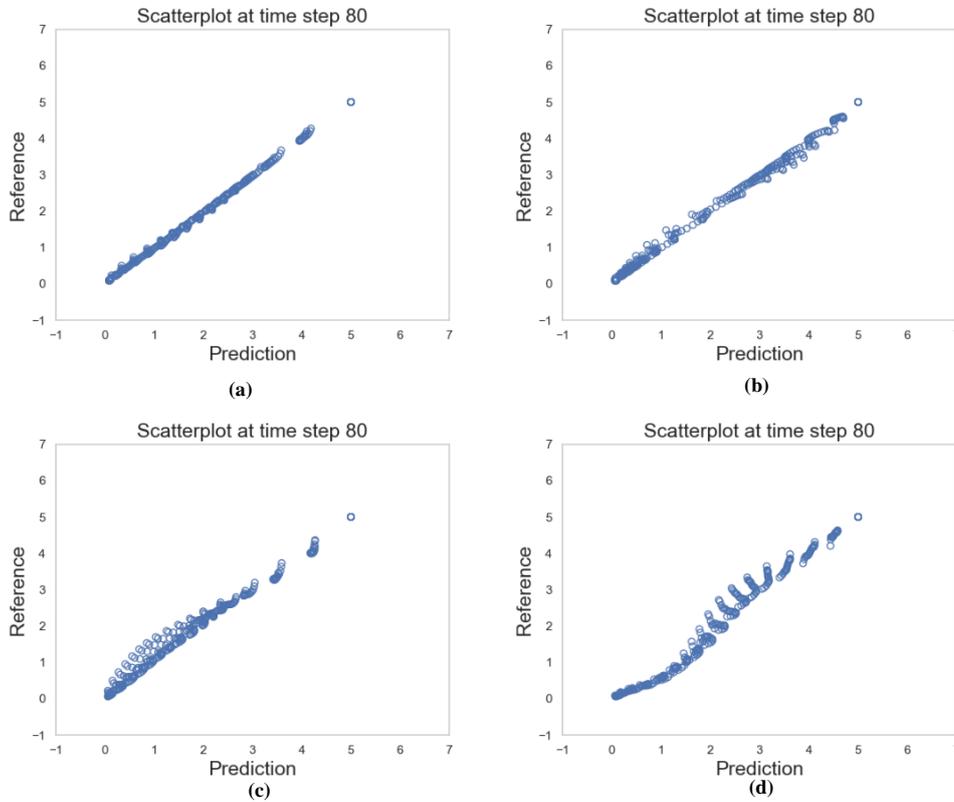

**Figure 14.** Scatterplots of the hydraulic head predicted by TgGAN for four hydraulic conductivity fields (a, b, c, and d).

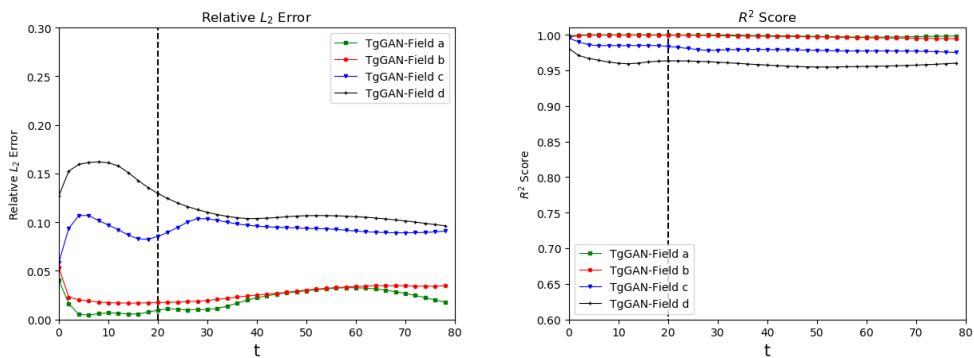

**Figure 15.** Changes of relative $L_2$ error and $R^2$ score over time for TgGAN for four hydraulic conductivity fields.

As can be seen from **Figure 15**, the results for field (a) are excellent since the TgGAN is trained with data from this particular field. Although TgGAN can make corresponding predictions for fields (b), (c) and (d), the accuracy is not stable. In other words, TgGAN does not show good generalization in this case, perhaps because the input field is a single one. Consequently, we next select 1, 5, and 10 hydraulic conductivity fields as training fields, respectively, to predict 50 new hydraulic conductivity fields to further investigate this issue. **Figure 16** shows the subset evaluations based on relative $L_2$ error and $R^2$ score. The running time for different numbers of training hydraulic conductivity fields is 2,234 s, 10,956 s, and 22,066 s, respectively, which increases nearly linearly with the number of training fields. It can be seen that, as the number of training hydraulic conductivity fields increases, the overall relative $L_2$ error is closer to 0, and the overall $R^2$ score is closer to 1, both of which have smaller standard deviations. Overall, the larger is the number of training $K$ fields, the stronger is the generalization ability of TgGAN.

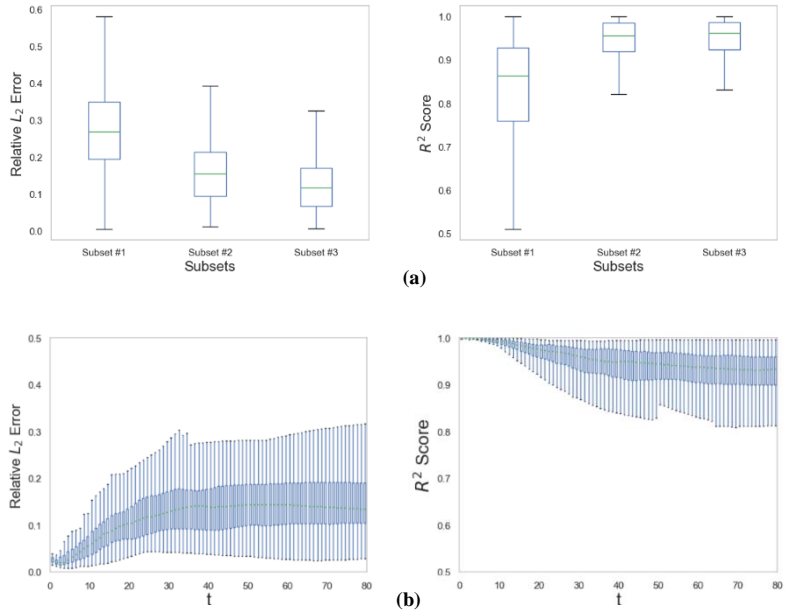

**Figure 16.** Subset evaluations based on relative $L_2$ error and $R^2$ score. (a): subset #1, #2, and #3 refer to cases in which the number of training hydraulic conductivity fields is 1, 5, and 10, respectively; (b): relative $L_2$ error and $R^2$ score over time for subset #3 .

### 3.7 Transfer learning

It can be seen in **subsection 3.6** that, although TgGAN performs better when increasing the number of training $K$ fields, the training time will increase correspondingly, which means that the performance of TgGAN is limited when facing complex problems. In order to improve the training efficiency of TgGAN by making use of the already trained TgGAN, transfer learning is employed (Pan and Yang, 2010).

In the transfer learning scheme, we firstly use simple conditions to train TgGAN

in the pre-training process, freeze the weights and bias of the last four convolutional layers, and then use complex conditions to train it again in the retraining process. The shallow layers of the network extract information about a particular system, while the deeper layers process the extracted information (Wang et al., 2020).

In the following two subsections, we aim to employ transfer learning to improve the performance of the case in **subsection 3.6**, and to predict the future response with changed boundary conditions.

### 3.7.1 *Predicting the corresponding hydraulic heads for new hydraulic conductivity fields*

During the pre-training process, one hydraulic conductivity field with variance of *ln K* being 0.5 is inputted into the TgGAN as a training field. 625 points are chosen randomly as training data for each of the first 20 time steps, amounting to a total of 12,500 data points (i.e., 625×20 time steps), and 2,209 points for each of 80 time steps are used as collocation points, amounting to a total of 176,720 collocation points (i.e., 2,209×80 time steps). Then, the weights and bias of the last four convolutional layers of generator are fixed, which have learned the information of physical information (Wang et al., 2020). In the transfer learning step, 5 and 10 hydraulic conductivity fields with variance of *ln K* being 1 in **subsection 3.6** are chosen as training fields, respectively, to train the generator independently to predict corresponding hydraulic heads for the other 50 new hydraulic conductivity fields with variance of *ln K* being 1. The retraining process, begins with the parameters of the pre-training generator, and no data points are used. The input dimensions of the generator are 400×3×51×51 and 800×3×51×51 for two different numbers of retraining fields (i.e., 5× and 10×80 time steps), respectively.

**Figure 17** shows relative $L_2$ error and $R^2$ score for the normal training (**subsection 3.6**) and transfer learning when the number of retraining fields is 5 and 10, respectively. The relative $L_2$ error and $R^2$ score over time when the number of retraining fields is 10 are shown in **Appendix A.4.1**. The running time of the pre-training process is 1,798 s. When the number of retraining fields is 5 and 10, the respective running time of transfer training processes is 3,491 s and 6,971 s. Compared to training directly with 5 and 10 fields reported in **subsection 3.6**, the employment of transfer learning can significantly reduce training cost without compromising accuracy.

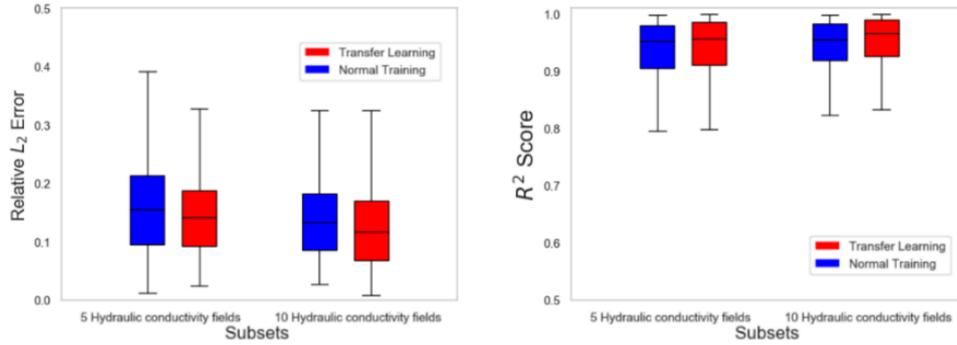

**Figure 17.** Relative $L_2$ error and $R^2$ score for normal training and transfer learning.

### 3.7.2 *Predicting the future response with changed boundary conditions*

In this case, we examine the ability of TgGAN for predicting the hydraulic head when the boundary condition is different from the training data. The training data are obtained with the following boundary condition: $H_{x=0} = 1[L]$ and $H_{x=1020} = 0[L]$ at the first 19 time steps. At the 20$^{th}$ time step, the right boundary condition is changed from 0 [L] to 2 [L]. In this case, the general flow direction is reversed when the boundary condition is changed. The input dimension of the generator is 70×3×51×51, where, 70 means the maximum number of time steps.

In normal training, we set two soft boundary constraints to train TgGAN. All data points at the first 19 time steps are used as training points, and 112,000 collocation points (i.e., 1,600 at each time step from 0 to 70) are used to predict the flow process of the following 50 steps. GAN is also trained with available data, but without collocation points.

In the transfer learning process, unlike normally trained TgGAN, the changed boundary condition in the last 50 time steps is assumed to be unknown beforehand. We aim to employ transfer learning to force TgGAN to learn the new boundary condition by making use of the pre-trained model. In the pre-training process, the TgGAN has been firstly trained using data and collocation points from the first 19 time steps with the following boundary condition: $H_{x=0} = 1[L]$ and $H_{x=1020} = 0[L]$. Then, the kernel weights and bias of the last four convolutional layers of the generator are fixed. In the transfer learning process, 80,000 collocation points (i.e., 1,600 at each time step from 20 to 70) are used to train the generator independently to examine the influence of the new boundary condition ($H_{x=0} = 1[L]$ and $H_{x=1020} = 2[L]$) by updating the parameters of the first two convolutional layers.

The predictions of normally trained TgGAN, GAN, and TgGAN trained by transfer learning at time step 70 are presented in **Figure 18**. Likewise, the changes of relative $L_2$ error and $R^2$ score over time for the three treatments are shown in **Figure 19**. The scatterplots and the distributions of the hydraulic head are shown in **Appendix A.4.2**. The running time of normally trained TgGAN, GAN, and TgGAN trained by transfer learning is 1,929 s, 1,910 s, and 1,183 s respectively. It is obvious that the training time of the new TgGAN model with changed boundary condition is significantly decreased by adopting transfer learning.

As can be seen from **Figure 19**, both the relative $L_2$ error and the $R^2$ score for GAN deteriorate from time step 20 since no data are available after the boundary condition changes. However, the quality of extrapolation from the normally trained TgGAN is still adequate owing to the fact that the change of boundary condition is reinforced via incorporating such information into the loss function. By employing the transfer learning technique, the quality of extrapolation is further improved, which means that the generator learns the new boundary condition in an efficient manner. When the conditions change in time, the strategy of transfer learning is recommended, as it leads to a sequential learning process, built upon what has been learned in the past.

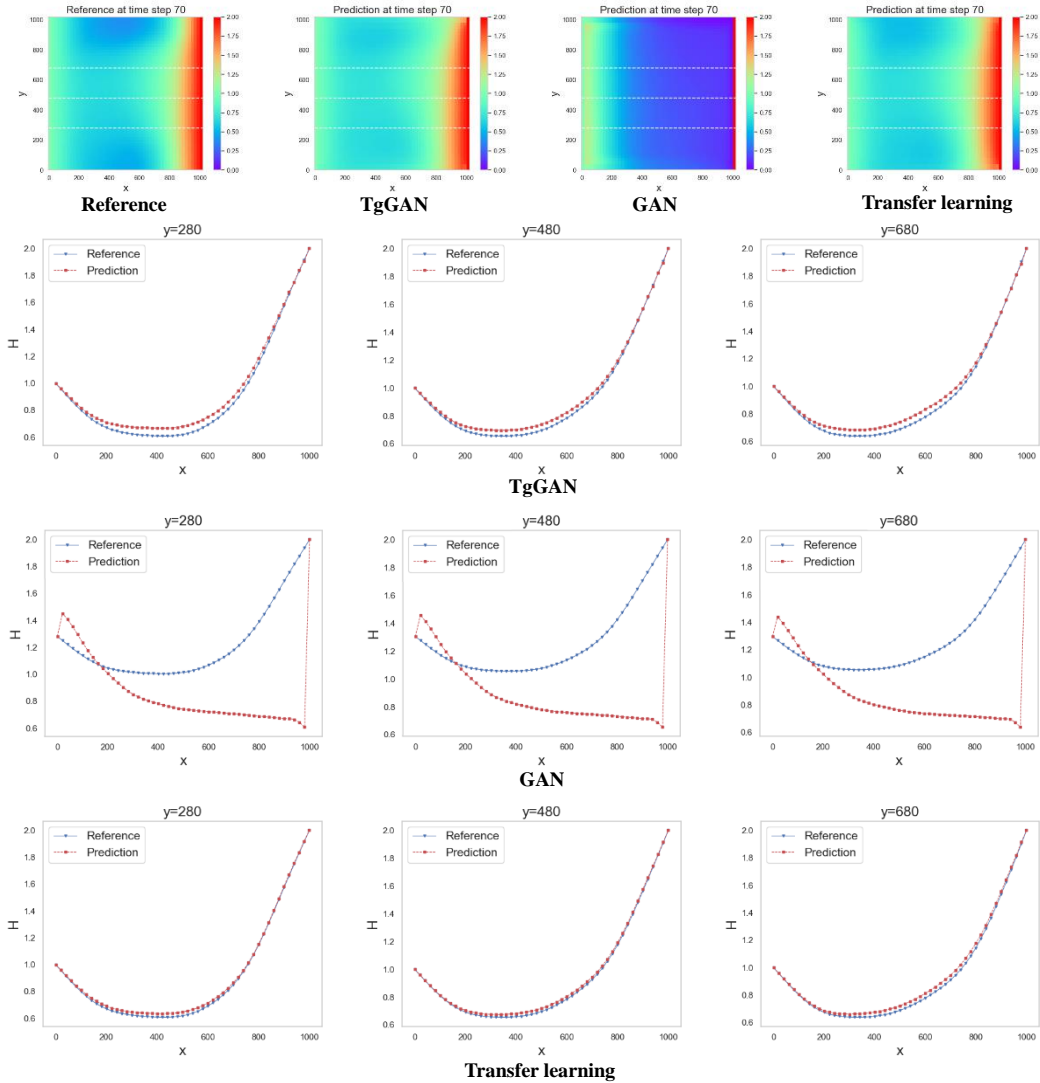

**Figure 18.** Predictions of the normally trained TgGAN, GAN, and TgGAN trained by transfer learning at time step 70.

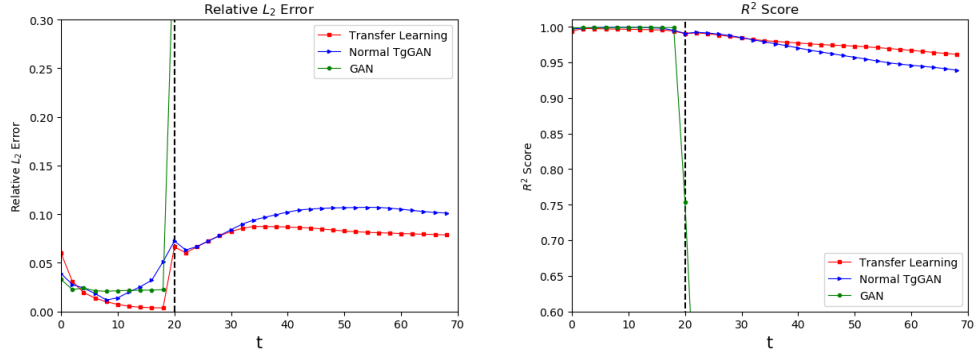

**Figure 19.** Changes of relative $L_2$ error and $R^2$ score over time for normally trained TgGAN, GAN, and TgGAN trained by transfer learning.

## 4. Discussion and Conclusions

In this study, we proposed the theory-guided generative adversarial network (TgGAN), which is a special framework for deep learning of dynamic subsurface flow. In TgGAN, we change the learning objective to the distribution of residuals between the training data and the generated data in order to accommodate the dynamic nature of the problem. In addition, certain theories (i.e., governing equations, boundary conditions, initial conditions, and other kinds of constraints) are incorporated into the loss function of the generator in the form of soft constraints, which can assist the TgGAN model to achieve better noise-resistance, stability, and robustness. Additional theories allow TgGAN to be more than just data-driven, and consequently enable the generator to perform excellent interpolation and extrapolation when training data points are scarce.

The performance of the proposed TgGAN is tested by several cases: predicting future response, predicting future response from noisy data, label-free learning or solving PDEs via deep learning, and predicting with changed boundary conditions. The effects of the number of training data points and collocation points are discussed, respectively. Compared with the standard GAN, TgGAN has superior stability with the enforcement of theories and can maintain excellent performance in several kinds of complex situations. The transfer learning technique is also employed to decrease the training time of TgGAN.

Similar to the standard GANs and other variants, TgGAN also has the problem of low learning efficiency. In addition, the weight of the constraint term in the loss function is challenging to balance. If the weight is too large, the loss of the training data is difficult to converge, resulting in an increase of computational cost. Conversely, if the weight is too small, the generator may ignore the theoretical constraints, and then extrapolation performance may diminish. Future studies may focus on improving the learning efficiency of TgGAN, and making use of the automatic differentiation technique to replace the numerical difference method to shorten the training time and further improve accuracy.

**Acknowledgements**

This work is partially funded by the National Natural Science Foundation of China (Grant No. 51520105005) and the National Science and Technology Major Project of China (Grant No. 2017ZX05009-005 and 2017ZX05049-003).

**Appendix A**

**A.1 Predicting the future response**

In this case, the hydraulic head distribution at the first 20 time steps is monitored, and 625 points are chosen randomly as training data for each of the first 20 time steps. At every other time step during the entire process of 80 time steps, 576 points are extracted as collocation points, amounting to a total of 23,040 (i.e., 576×40 time steps) collocation points. Both the collocation points and the data points are randomly selected in space, but the positions of these points are fixed in time. The distribution at the following 60 times steps is expected to be extrapolated. The predictions from TgGAN and GAN at time step 80 with hydraulic conductivity fields (b), (c), and (d) are shown in **Figure A.1**.

**Field (b)**

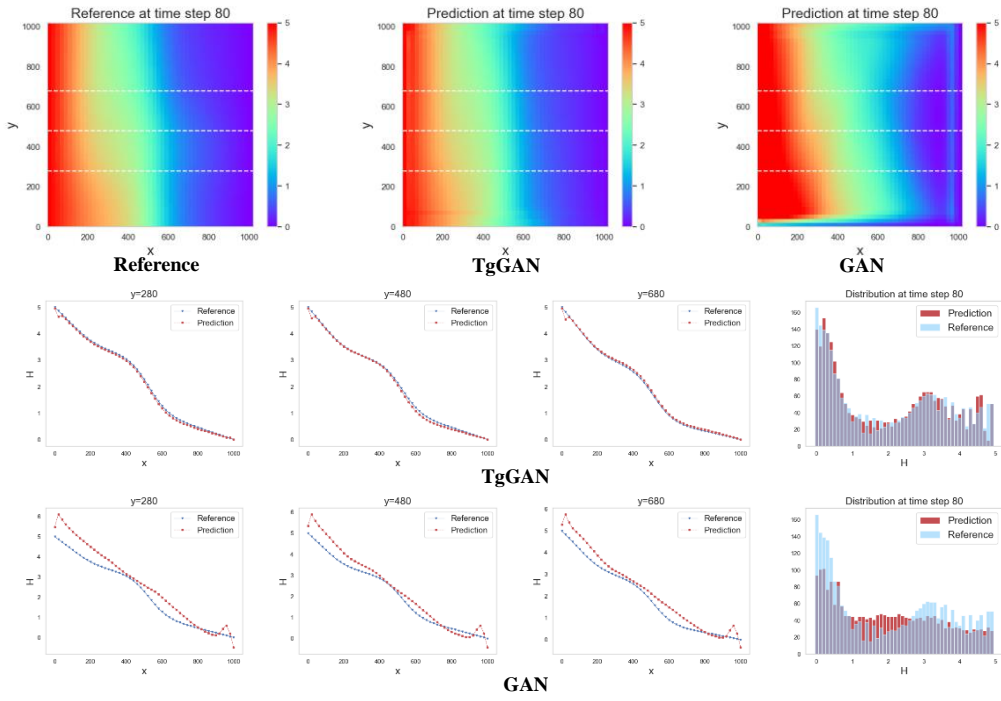

**Field (c)**

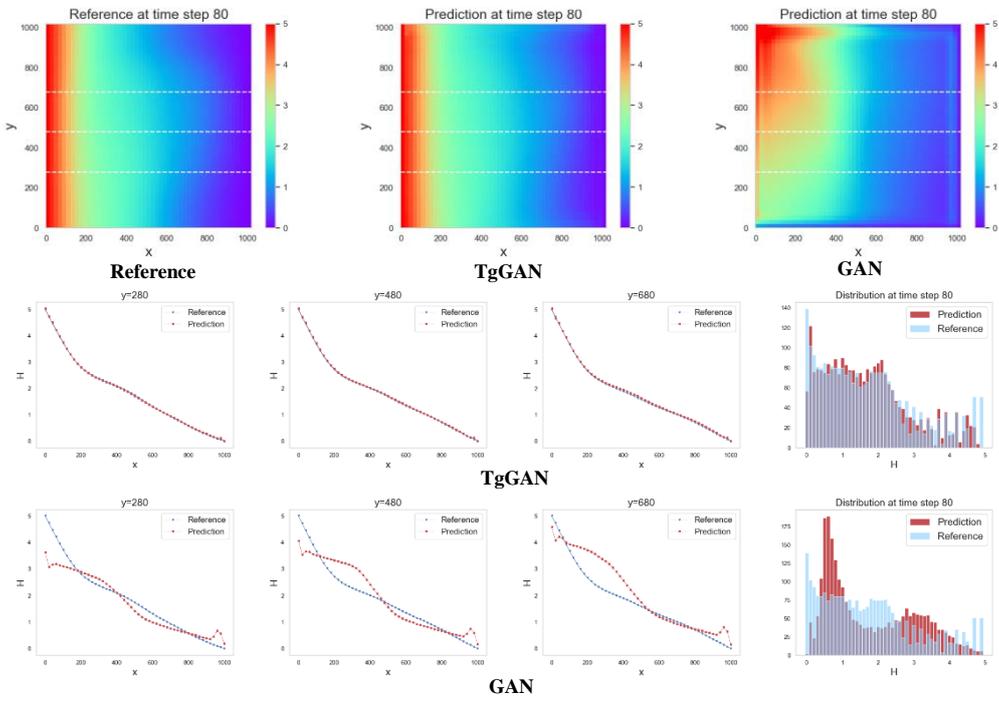

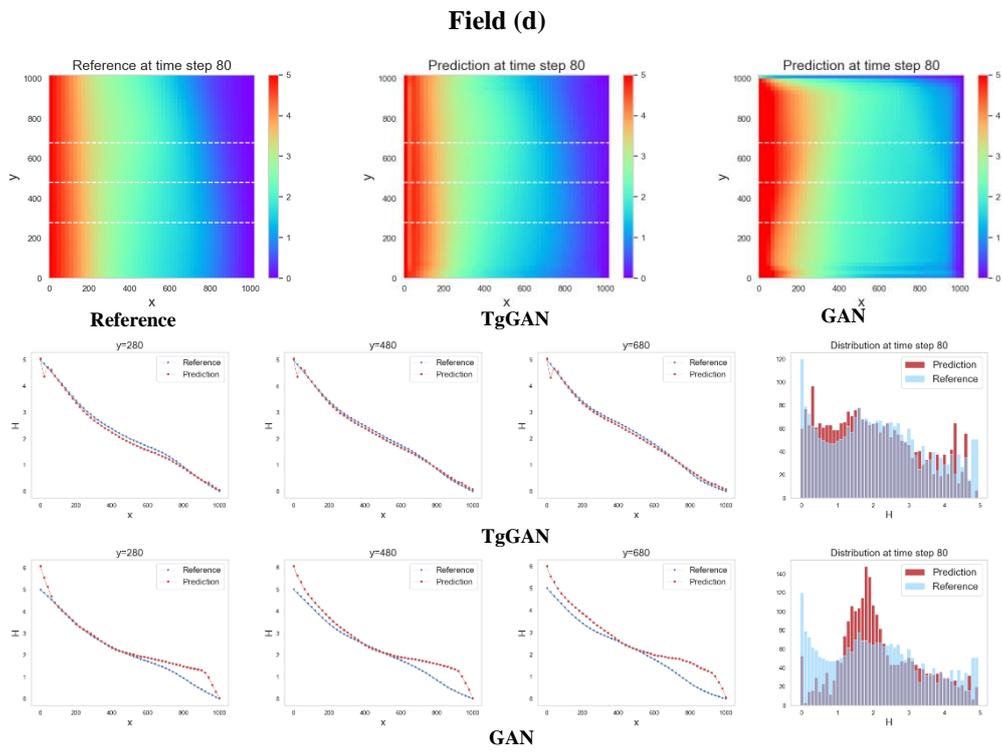

**Figure A.1.** Predictions of TgGAN and GAN for hydraulic conductivity fields (b), (c) and (d).

## A.2 Predicting the future response from noisy data

In this case, we test the ability of TgGAN and GAN to resist different levels of noise. The results of TgGAN and GAN at time step 80 are shown in **Figure A.2.1**. The head distributions for TgGAN and GAN at time step 80 when noise exists are shown in **Figure A.2.2**. It can be seen that TgGAN and GAN have different noise resistance to various degrees, and TgGAN is significantly more stable.

**(5% noise)**

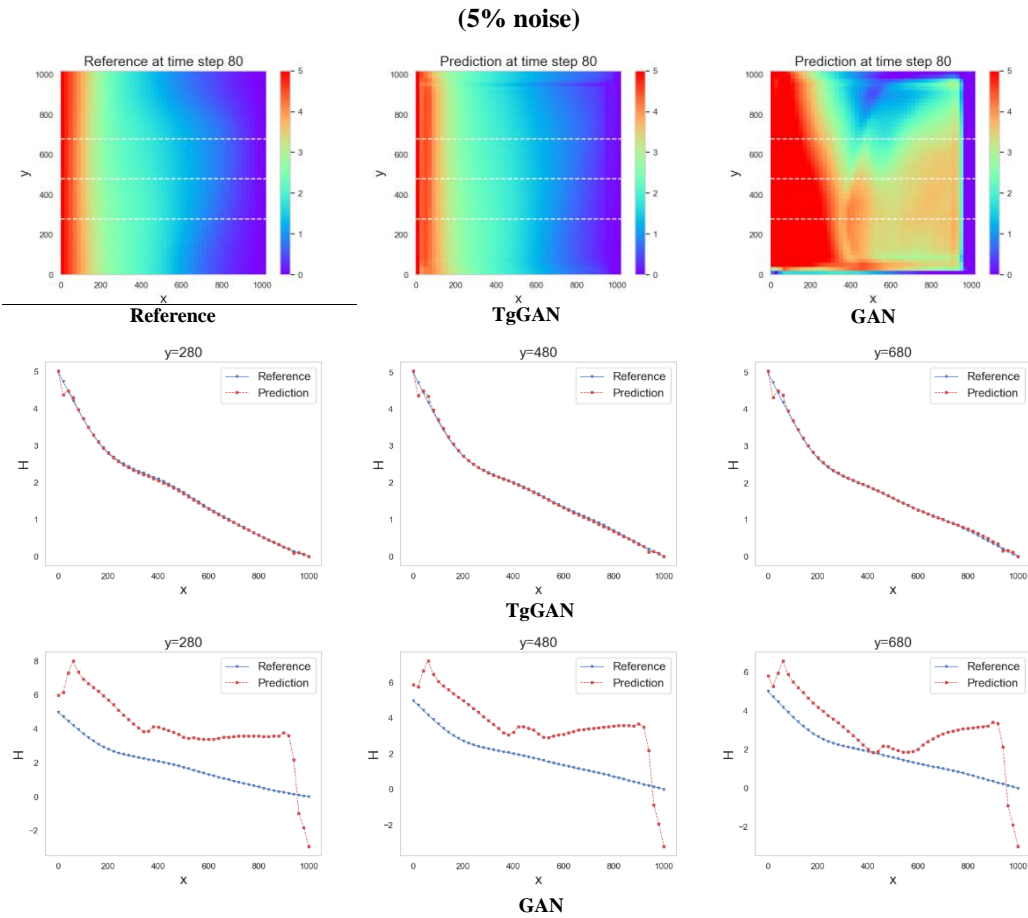

**(10% noise)**

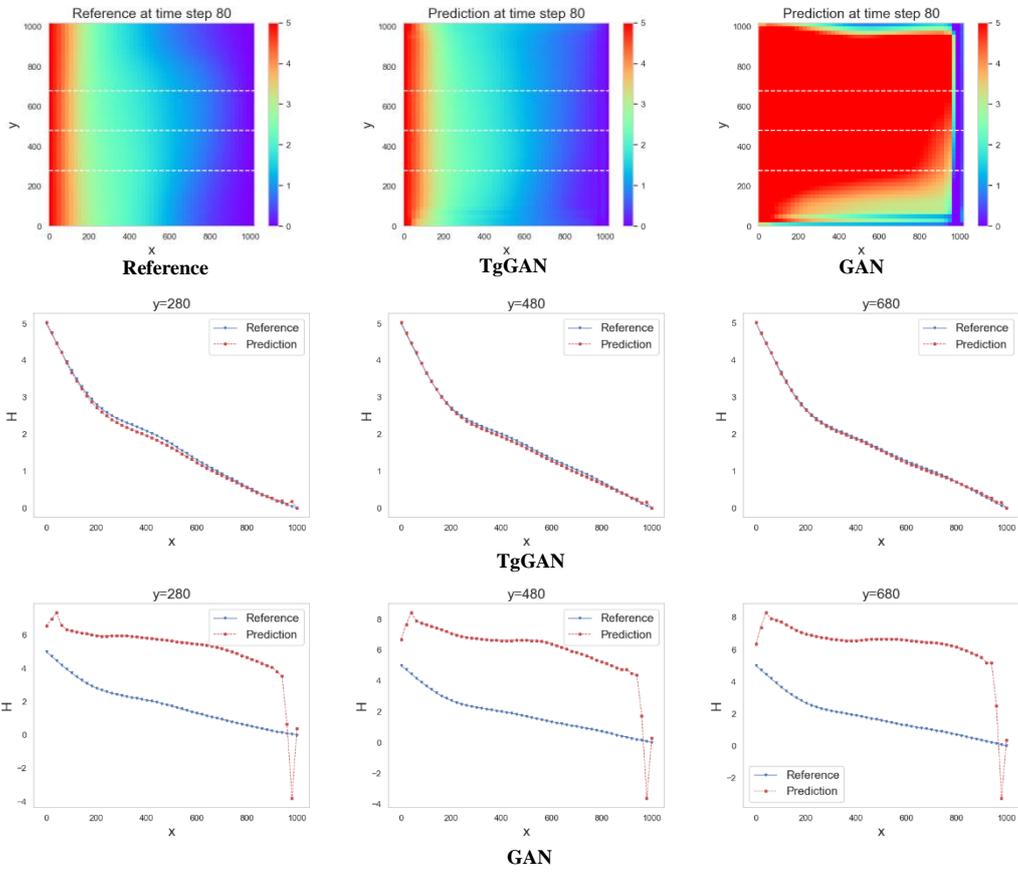

**(20% noise)**

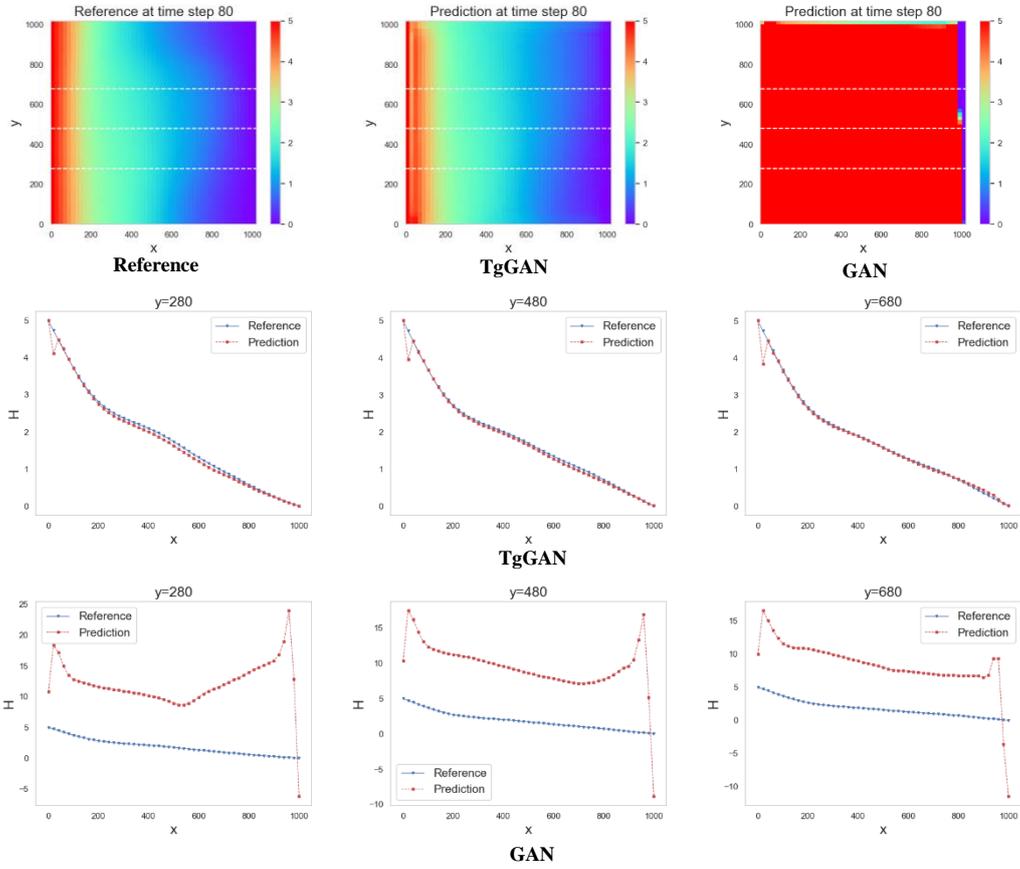

**Figure A.2.1** Predictions of TgGAN and GAN trained from noisy data.

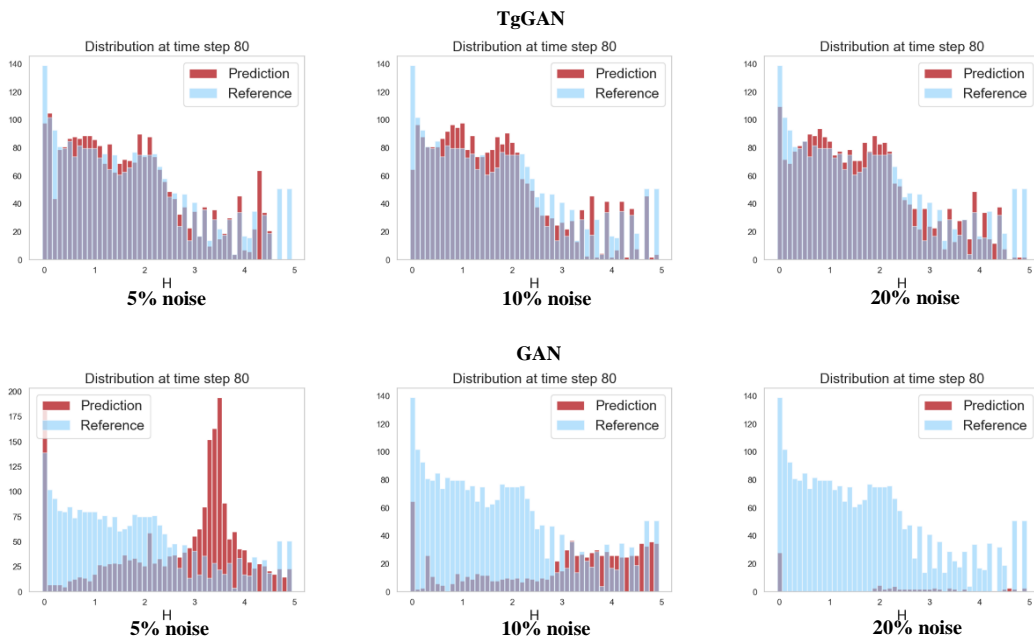

**Figure A.2.2** Head distributions for TgGAN and GAN at time step 80 when noise exists.

## A.3 Predicting the future response without labels

In this case, we aim to test the ability of label-free learning of solving PDEs via TgGAN. The predicted results at time step 30, 50 and 70 are shown in the following figures.

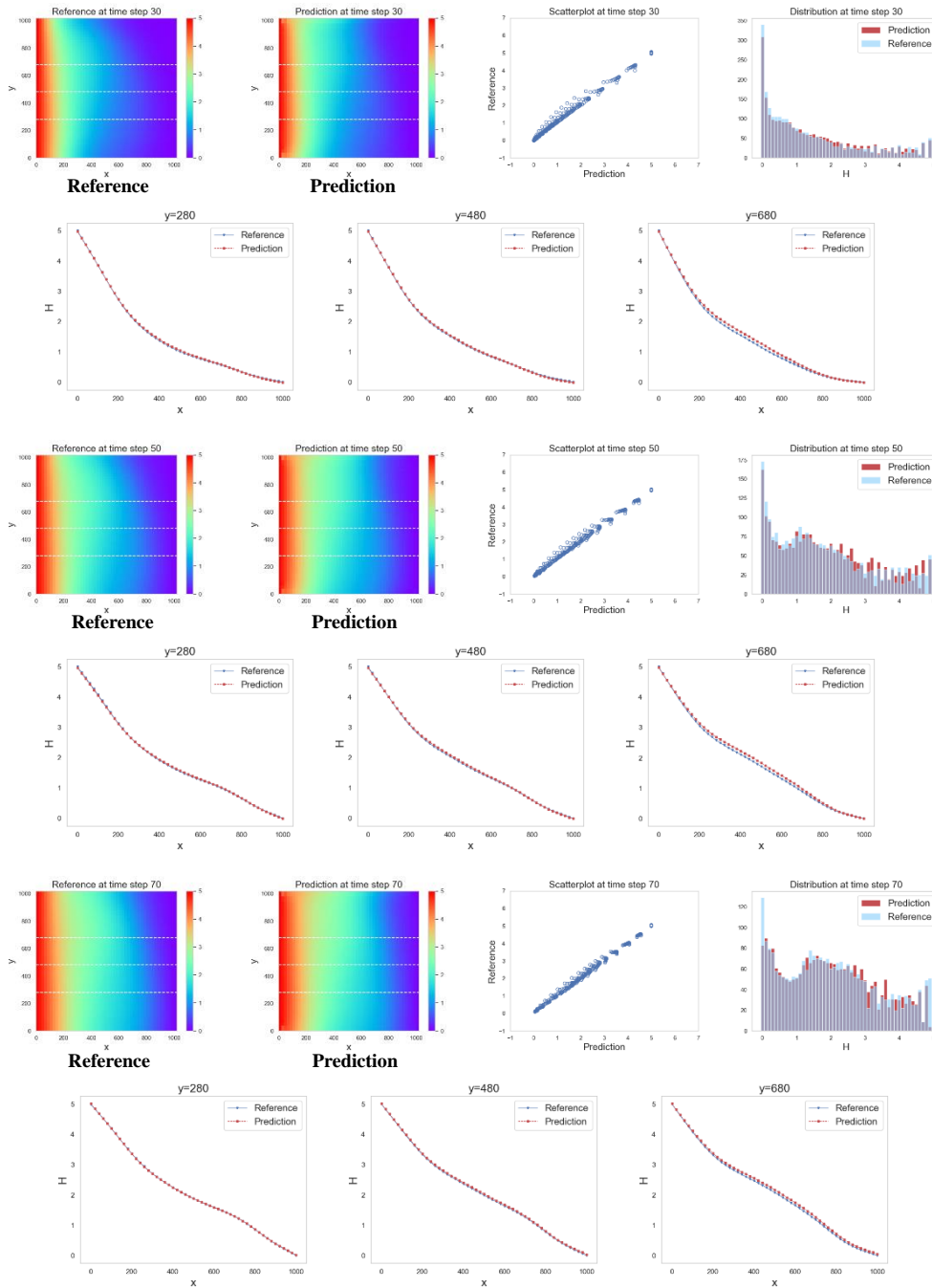

**Figure A.3** Predictions of the hydraulic head at different time steps (30, 50, and 70) via TgGAN without labels.

## A.4 Transfer learning

In theory, as the number of training hydraulic conductivity fields increases, the generalization ability of TgGAN will become stronger, while the training time and computational cost will increase correspondingly.

In this subsection, in order to improve the training efficiency of TgGAN, transfer learning is employed. **Figure A.4.1** shows subset evaluations based on relative $L_2$ error and $R^2$ score over time when the number of retraining fields is 10. Then, the transfer learning technique is also employed to predict future responses with changed boundary conditions. **Figure A.4.2** shows the scatterplots and the distributions of the hydraulic head at time step 70.

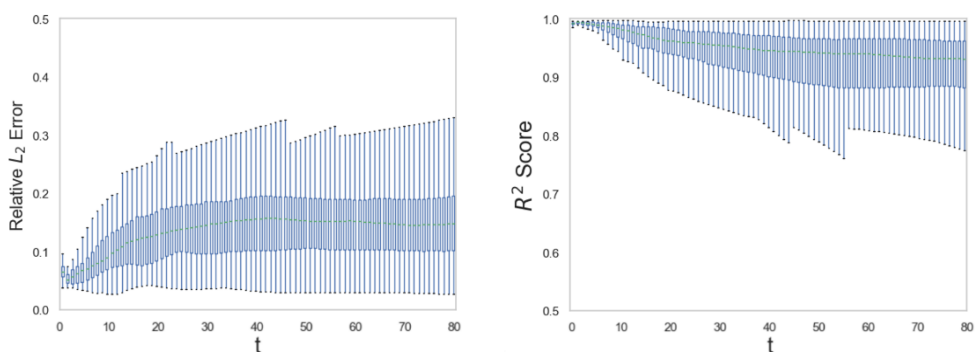

**A.4.1** Relative $L_2$ error and $R^2$ score over time when the number of retraining fields is 10.

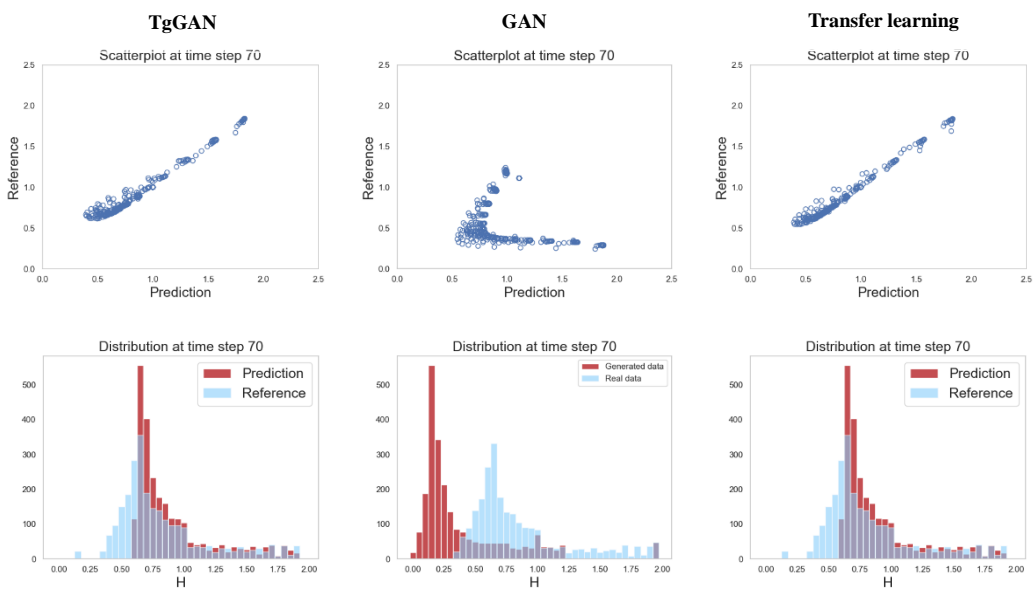

**A.4.2** Scatterplots and distributions of the hydraulic head at time step 70.